  \pdfoutput=1
  \documentclass[graphicx,amssymb,amsmath,enumerate,11pt]{article}
  \usepackage{fancyhdr}
  \usepackage{subfigure}
    \usepackage{url}
\usepackage{amssymb}
\usepackage{amsmath}
\usepackage{amsfonts}
\usepackage{array,hhline,dcolumn} % Better table handling
  \fancyheadoffset[]{0.1cm}
\usepackage[sort&compress,numbers]{natbib}
  \usepackage{epsfig}
  \usepackage{url}
  
\usepackage{suffix}
\usepackage{color}
\usepackage{color}
\usepackage{bm}
\usepackage{tikz}

\newcommand{\INLINEBOX}[2]{%
   \begin{center}%
    \fcolorbox{#1!60!black}{#1}{%
      \addtolength{\linewidth}{-0.8cm}%  fixed value, works for normal article text
      \begin{minipage}{\linewidth} #2 \end{minipage}%
    }%
   \end{center}\vspace{4pt}%
}

\newcommand{\MARGINBOX}[1]{%
  \mbox{}%
  \marginpar%
   [\tiny\raggedleft\hspace{0pt}#1]%
   {\tiny\raggedright\hspace{3pt}#1}%
}

\newcommand{\TODO}[2][]{\MARGINBOX{\textcolor{red!80!black}{\emph{ToDo (#1):}} #2}}
\WithSuffix\newcommand\TODO*[2][]{\INLINEBOX{red!20!white}{\emph{ToDo (#1):} #2}}

\newcommand{\FIXME}[2][]{\MARGINBOX{\textcolor{blue!80!red}{\emph{FixMe (#1):}} #2}}
\WithSuffix\newcommand\FIXME*[2][]{\INLINEBOX{blue!20!red}{\emph{FixMe (#1):} #2}}

\newcommand{\NOTE}[2][]{\MARGINBOX{\textcolor{green!80!black}{\emph{Note (#1):}} #2}}
\WithSuffix\newcommand\NOTE*[2][]{\INLINEBOX{green!20!white}{\emph{Note (#1):} #2}}

\newcommand{\opal}{\textsc{OPAL}}
\newcommand {\htp}{$\text{H}_2^+$}
\newcommand {\hmi}{$\text{H}^{-}$}
\begin{document}      %!%   Comment this line  with %!%

\begin{center}
{\Large \bf  Cyclotrons as Drivers for Precision Neutrino Measurements}\\

{\it A.~Adelmann~\footnote{Paul Scherrer Institut, Villigen, CH-5232, Switzerland}, J.~Alonso~\footnote{Massachusetts Institute of Technology, Cambridge, MA 02139, USA}, W.A. Barletta~$^2$,
  J.M.~Conrad~$^2$, M.H.~Shaevitz~\footnote{Columbia University, New York, NY 10027, USA}, J.~Spitz~$^2$, M.~Toups~$^2$, and
  L.A.~Winslow~\footnote{University of California, Los Angeles, Los Angeles, California 90095, USA}}
\end{center}

\vskip 0.5cm

\noindent{\bf Abstract:}\\
{\small As we enter the age of precision measurement in neutrino
  physics, improved flux sources are required.  These must have a
  well-defined flavor content with energies in ranges where
  backgrounds are low and cross section knowledge is
  high.   Very few sources of neutrinos can meet these requirements. However, pion/muon and isotope decay-at-rest sources qualify. The ideal
  drivers for decay-at-rest sources are cyclotron accelerators, which are compact and
  relatively inexpensive.   This paper describes a scheme to produce decay-at-rest sources driven by such cyclotrons, developed within the
  DAE$\delta$ALUS program.   Examples of the value of the high
  precision beams for pursuing Beyond Standard Model interactions are reviewed.   New
  results on a combined DAE$\delta$ALUS--Hyper-K search for
  $CP$-violation that achieve errors on the mixing matrix parameter
  of 4$^\circ$ to 12$^\circ$ are presented.}

\vskip 1cm

\section{Introduction}
As we reach the 100th anniversary of the birth of Bruno Pontecorvo,
neutrino physics is facing a transition.    Neutrino
oscillations are well established, albeit in a different form from
what Pontecorvo expected~\cite{pontecorvo1, pontecorvo2}.   We have a data-driven ``Neutrino
Standard Model,'' ($\nu$SM) which, despite questions
about its underlying theoretical description, is remarkably predictive.  Now, the neutrino community must pivot from ``searches''
to ``precision-measurements,'' in which we can test the $\nu$SM.    
The transition requires new and better tools for these measurements and further calls for original approaches to experiments.

\begin{figure}[t]
\begin{center}
\includegraphics[scale=.4]{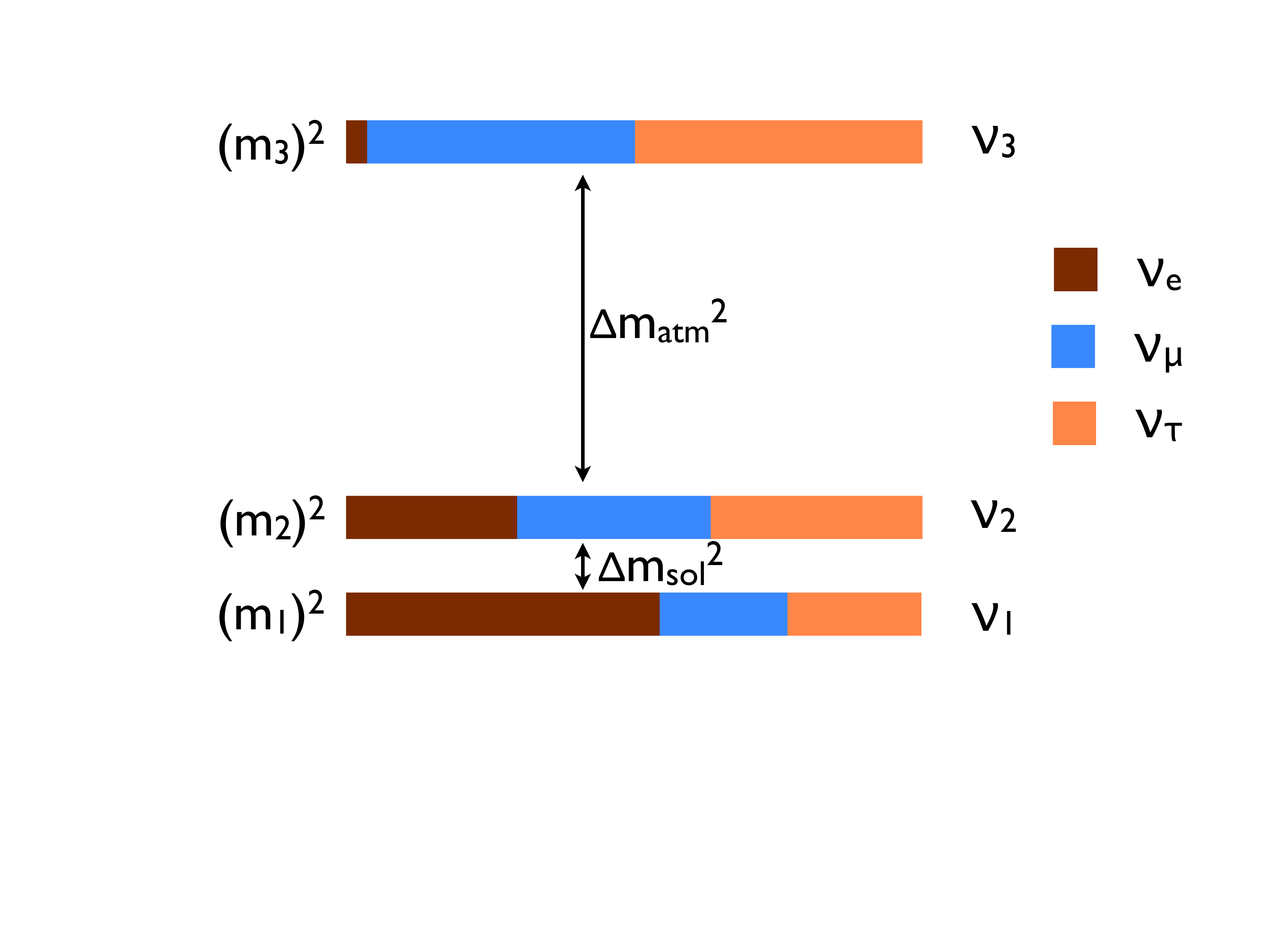}
\end{center}
\vspace{-3.5cm}
\caption{Illustration of the ``$\nu$SM'' showing mass states and
  mixings.  Note that this drawing depicts only one possible
  mass-ordering. There remain many open questions that surround this data-driven picture of neutrinos and oscillations.  
\label{nuSM}}
\end{figure}

The $\nu$SM is simply described in Figure~\ref{nuSM}.  The three known neutrino flavors mix within three mass states.  The
separations between the states, or ``mass splittings,'' are 
defined as $\Delta m^2=m_j^2-m_i^2$, for $i,j=1 \dots 3$.    The
historical name for the smaller splitting ($\Delta m_{21}^2$) is $\Delta
m_{sol}^2$ and the larger mass splitting ($\Delta m_{31}^2 \approx
\Delta m_{32}^2)$ is referred to as $\Delta m_{atm}^2$, in honor of the solar and
atmospheric experiments that established the existence of each.  The early solar~\cite{chlorine, SAGE, GALLEX, Kamsol} and
atmospheric~\cite{Kamatm, IMB, Soudan} experiments have been joined by
new results~\cite{SKsol, SNO, KamLAND, K2K, MINOS, T2K} to establish 
this phenomenology~\cite{PDG}.   

The mixings are described with a 3$\times$3 matrix, commonly called the Pontecorvo-Maki-Nakagawa-
Sakata (PMNS) matrix, 
 that connects the mass
eigenstates ($\nu_1$, $\nu_2$, $\nu_3$) to the flavor eigenstates 
($\nu_e$, $\nu_\mu$, $\nu_\tau$) :
\begin{equation}
U =
\begin{pmatrix}
 U_{e 1 } & U_{e 2}  & U_{e 3}   \cr
 U_{\mu 1  } & U_{\mu 2}  & U_{ \mu 3}   \cr
 U_{\tau 1  } & U_{\tau 2}  & U_{ \tau 3}   
\end{pmatrix} 
=
\begin{pmatrix}
0.795-0.846 & 0.513-0.585 & 0.126-0.178  \cr
0.205-0.543 & 0.416-0.730  & 0.579-0.808 \cr
0.215-0.548 & 0.409-0.725 & 0.567-0.800 
\end{pmatrix}, 
\end{equation}
where the ranges indicate our knowledge
of each of the entries \cite{PMNSranges}. Together with the mass splittings, the mixing matrix is
pictorially represented in Figure~\ref{nuSM},  in which
the lengths of the
colored bars are proportional to the squared moduli of
the matrix elements, $|U_{\alpha i}|^2$.

As can be seen,  our current knowledge of the mixing matrix
values is imprecise.
The last entry, $U_{e3}$, was found to be non-zero just two
years ago~\cite{DC, DB, RENO, T2Kapp}.   Our current state
of measurement of the neutrino mixing matrix is analogous to that 
of the quark sector in 1995~\cite{pdg1995}, immediately after 
the discovery of the top quark.
Unlike in the quark sector and its utilization of strong production, 
neutrino physicists are faced with the difficulty of both weak production and weak
decay. Our route to precision therefore drives us  to 
high-intensity sources coupled with ultra-large detectors.

Even at this relatively early stage, the $\nu$SM has been remarkably predictive. For example, the
$|U_{e3}|$ element was found with the $\Delta m^2_{atm}$ splitting~\cite{Thiago} as expected from the model.  However,  many open questions remain.   Figure~\ref{nuSM}
shows a ``hierarchy'' of the mass states, after arranging the large and small splittings
so that the orientation is consistent with what is seen in the quark sector.    It is unclear if the neutrinos are oriented in a ``normal hierarchy'', as shown, or if the orientation is actually ``inverted''. Further, although the values of each of the splittings have been measured, the absolute mass of the neutrino is not known. We know there is a $3\times3$ matrix that describes the mixing but we don't know if there is $CP$-violation present as in the quark sector. There are also more exotic questions surrounding the neutrino and oscillations. For example, are there new
forces appearing in neutrino interactions and oscillations? Do
exotic non-interacting (``sterile'') neutrinos mix with the known
active flavors?   Hints for all of these possibilities exist~\cite{hierarchy2013, massscale2013, delta2013, nsi2013, sterile2013} with evidence extending up to $4\sigma$.  The next generation of neutrino
experiments must investigate these results and clarify the present
picture. If history is any indicator of the future, it is quite likely that these experiments, along with the more conventional ones within the $\nu$SM, will raise even more surprises.  

The next step in neutrino physics requires improved tools, in
particular, sources from which the energy distribution and flavor
content are very well-defined.  The beam energy must be in a range where the
neutrino interaction cross section is understood, backgrounds are low, and where the detectors are
highly efficient.  Decay-at-rest (DAR) sources satisfy these requirements and provide an opportunity for the precision measurements required for the future of neutrino physics.

This paper explores cyclotrons as a relatively low cost means of
producing DAR sources at or near
underground laboratories.  We begin by discussing the pros and cons of
DAR for neutrino physics.   Next, we review the history of 
and development of cyclotrons.  
We then describe the machines under development within the
DAE$\delta$ALUS program. 
The final sections of this paper provide examples of
the precision science opened up by the DAR sources.   We explore tests in neutrino oscillations and
neutrino scattering.    We also discuss the potential impact beyond
particle physics.

\section{Decay-at-rest Sources of Neutrinos}
The most common source presently used in neutrino physics is  
the ``conventional muon-neutrino beam.''  Such a source is 
produced with GeV-scale protons striking a target, resulting in
pions and kaons which decay-in-flight (DIF) to produce neutrinos. The energy
distribution and relevant backgrounds of a given DIF beam are dependent on the design of the beamline. For example, the characteristics of a beam are quite sensitive to the magnetic focusing and decay region geometry as well as primary/secondary hadron production and interaction physics in the target. These complications make first principles predictions of the flavor-dependent neutrino energy distributions for both signal and background from DIF beams quite difficult.  

A number of techniques are available for DIF-based experiments to
understand the neutrino flux. Experiments with very high interaction
rates can use data to constrain the flux. For example,  the MiniBooNE
experiment has successfully used $\nu_\mu$ interactions, which come
largely via pion DIF from the Booster Neutrino Beamline (BNB) at Fermi National Laboratory, to constrain the
$\nu_e$ backgrounds that are due to the subsequent decay of the pion's daughter muon~\cite{MBFlux}.  Long baseline experiments use both a near detector and a far detector to reduce flux uncertainties.  This comparison works well for charged current interactions, in which the neutrino energy can be fully reconstructed. Neutrinos
which are within the acceptance of the far detector are considered in
measuring the flux of the near detector. However, this approach is not possible for
neutral current background events, because the
neutrino energy cannot be fully reconstructed.   Lastly, a wide range of event topologies are produced by conventional
beams, which range in energy from hundreds of MeV (e.g. JPARC~\cite{JPARC}, BNB~\cite{MBFlux}) up to  tens of GeV (e.g. NuMI~\cite{NuMI},
CNGS~\cite{CNGS}).  The cross sections for neutral current events and topologies involving pion production are not well
measured and understood~\cite{FormaggioZeller} and the pions (and their decay products) produced in the events can lead to electron-like backgrounds.

\begin{figure}[t]
\begin{center}
\includegraphics[scale=.25]{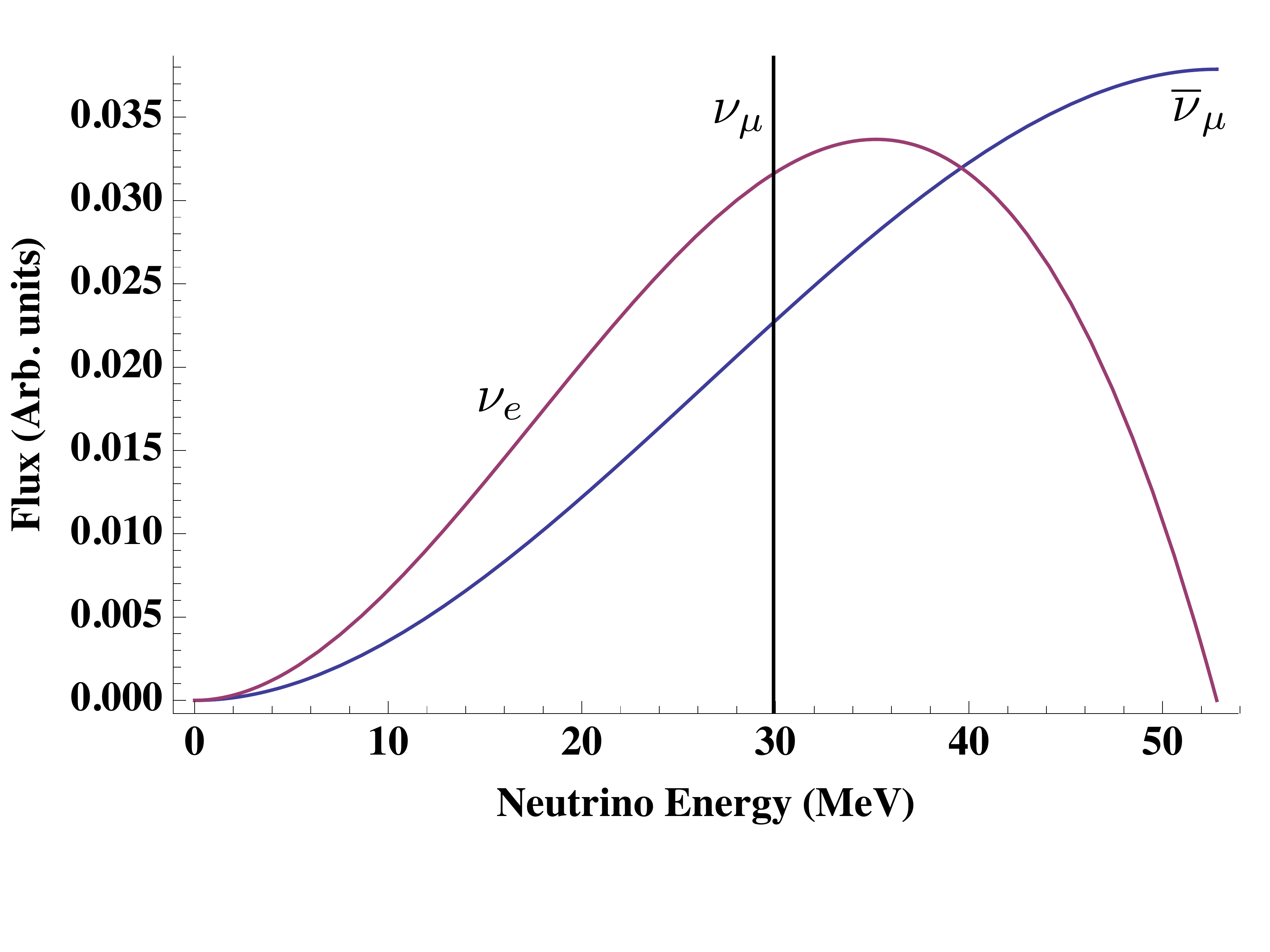}
\end{center}
\vspace{-1.5cm}
\caption{Neutrino flux distribution from a pion/muon DAR source, from Ref~\cite{Anderson:2012pn}.}
\label{flux}
\end{figure}

DAR neutrino sources, produced through $\pi \rightarrow \mu
\rightarrow \nu$ and isotope decay, offer a precision alternative to DIF beams. DAR flux sources have well defined flavor content and energy distributions, as can be seen with the flux from a $\pi \rightarrow \mu
\rightarrow \nu$ source shown in Figure~\ref{flux}.  However, the 
neutrino energy is low compared to DIF--ranging from a few MeV to 52.8~MeV.  The low energy of a DAR source is both an advantage and a disadvantage.   A great advantage is that two of the
best-known neutrino cross sections, each with less than 1\% uncertainty,  are accessible at DAR beam
energies. The first is the inverse beta decay interaction (IBD, $\bar
\nu_e + p \rightarrow e^+ +  n$).  This has a cross section that is well
known because it is connected to neutron decay, which is measured very
precisely~\cite{GrattaIBDxsec}.   IBD can be efficiently observed by requiring a coincidence between the prompt positron
and delayed neutron capture signal~\cite{KamLAND, DC, DB, RENO}. The coincidence signature also allows the signal to be easily distinguished from background, especially if the background is low as is the case for an underground detector.
The second is neutrino-electron elastic scattering ($\nu + e
\rightarrow \nu + e$).   This cross section is well constrained by Standard Model measurements of $e^+e^-$ scattering~\cite{nuepapers}.
Although this interaction lacks a coincidence signal,  it is highly directional, even at DAR energies.  On the other hand, 
the low energy neutrinos from a DAR source means that the relevant interactions have low absolute cross
sections, leading to the high flux requirement. A DAR source therefore  has the overall disadvantage of requiring a very intense source that can be installed at or near a detector in an underground location.    Below,
we will show that cyclotrons, as DAR neutrino drivers, have sufficiently high intensity and small enough size to overcome these disadvantages.

\begin{figure}[t]
\begin{center}
\begin{tabular}{c c}
\includegraphics[scale=.4]{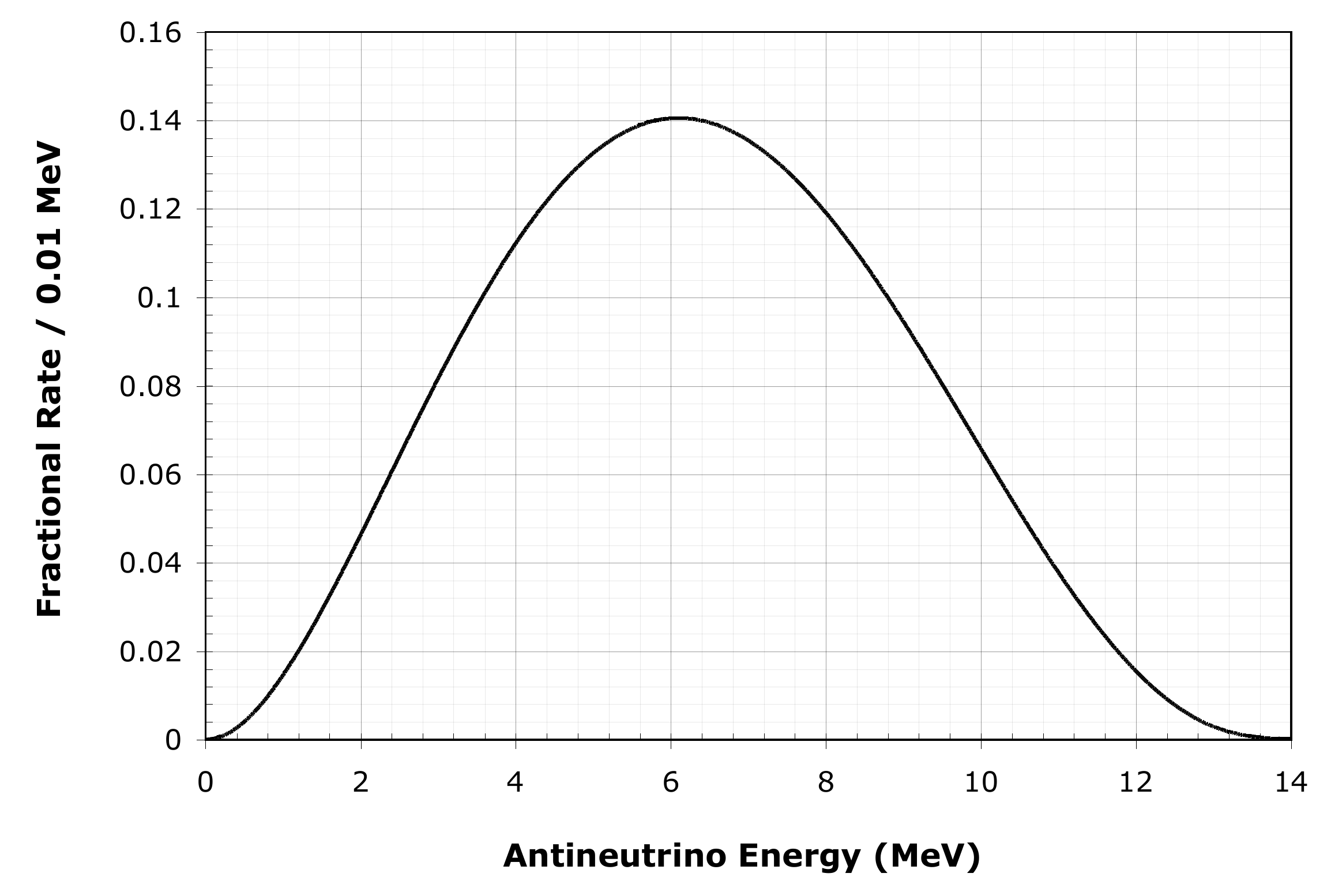}
\end{tabular}
\end{center}
\vspace{-.5cm}
\caption{The $^8$Li isotope DAR anti-electron-neutrino flux.}
\label{liflux}
\end{figure}

\begin{figure}[t]
\begin{center}
\begin{tabular}{c c}
\includegraphics[scale=.23]{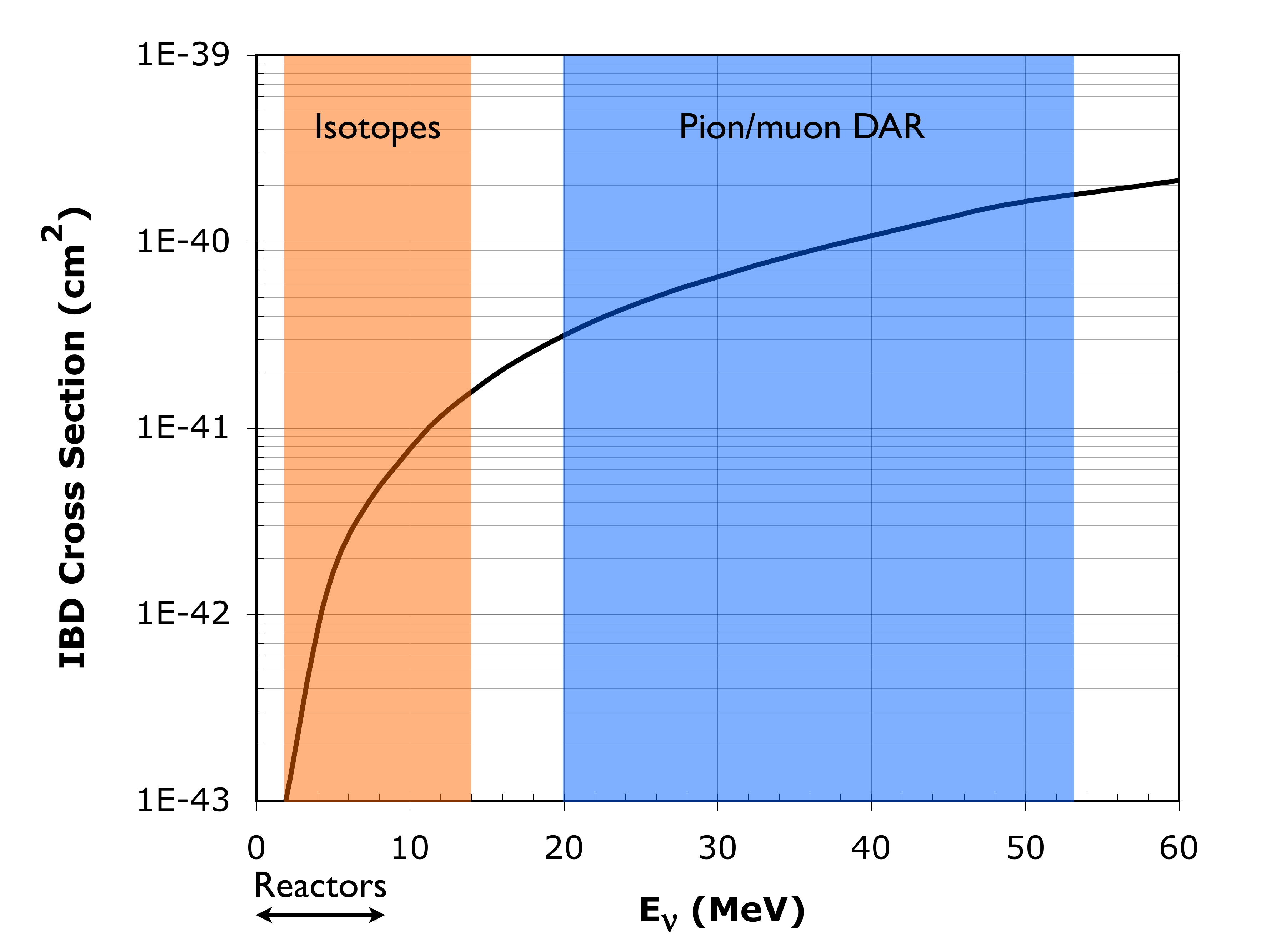}
\end{tabular}
\end{center}
\vspace{-.5cm}
\caption{Overlay of high flux regions on the IBD cross section.}
\label{flux_xsec}
\end{figure}

DAR sources range in energy from up to a few MeV
from isotope decay, where we use $^8$Li decay as our example (see
Figure~\ref{liflux}),  to 52.8~MeV from the $\pi^+ \rightarrow \mu^+$
chain (see Figure~\ref{flux}).  The flux from isotope decay is pure in flavor, while the
pion/muon decay has well-defined flavor ratios. 

The pion/muon DAR beam is best produced by impinging low energy
$\sim 800$~MeV protons on a light target to produce a high rate of
pions through the $\Delta$-resonance.  The target
must be surrounded by heavy material to stop the outgoing pions before DIF.
In this case, the neutrinos originate primarily from $\pi^+$ or
$\mu^+$ decay. The negatively charged pions and muons stop and capture 
on nuclei before they can decay to produce neutrinos~\cite{burman}.
The production of kaons or heavier mesons, which could produce unwanted
backgrounds, is negligible if the primary
proton energy is below about 1~GeV.
The $\bar \nu_e$ flux can be maintained at the level $\sim 5 \times 10^{-4}$ of the $\bar \nu_{\mu}$ flux
in the $20 < E_{\nu} < 52.8$ MeV energy range.  As a result,  the
source is well suited to
search for $\bar \nu_e$ appearance through oscillations~\cite{firstpaper}, as discussed below.  

An isotope DAR source produces a pure electron-flavor flux through 
$\beta$-decay. Such a source can be produced through high-intensity, low-energy
protons impinging on a beryllium target. These and subsequent interactions result in a flood of neutrons that
are captured on surrounding material to produce the isotope of
interest.  Precision experiments are best performed using neutrinos above 3 MeV,
where environmental backgrounds are low.   As a result, high $Q$-value
isotopes are favored.   Below, we discuss the use of $^8$Li as the decaying
isotope, produced by neutron capture in a 99.99\% pure $^7$Li sleeve surrounding the target.
This process produces a very pure $\bar \nu_e$ flux of well-defined energy
that can be used for scattering and neutrino disappearance experiments.

The IBD interaction has a large cross section at these MeV-scale energies, as shown in Figure~\ref{flux_xsec}. We note, however, that as the IBD interaction requires free
protons as an interaction target, this approach is relevant for water- and scintillator-based detectors only. The IBD interaction has a distinguished history. Pontecorvo himself first suggested a search for this interaction, in a 1946 report to the National Research Council of Canada~\cite{pontecorvoIBD, pontecorvoIBD2}, and it was the first type of neutrino interaction observed~\cite{discovery}. IBD is the signal interaction now widely used in reactor experiments. The reactor IBD range is shown in Figure~\ref{flux_xsec} in comparison to the DAR fluxes.

\section{Cyclotrons as DAR Source Drivers}
Cyclotrons represent ideal drivers for the DAR sources discussed above.
These machines are compact and low-cost compared to most particle physics accelerators.  The size, power, and cooling
requirements are sufficiently modest that it is possible to install cyclotrons at underground laboratories where these sources can be paired with existing large scintillator and water detectors.   Cyclotrons date back to Pontecorvo's era. However, and as in the case of neutrino physics, cyclotrons have come a long way since their origin. Modern cyclotrons are capable of producing the very high intensity flux required for modern precision neutrino measurements.

This section reviews the history of cyclotrons, with some discussion about how cyclotrons work.   We consider important
examples in use today and then discuss their development within the DAE$\delta$ALUS project.

\subsection{A brief history of cyclotrons}
Unbeknownst to Ernest O. Lawrence, the cyclotron was first invented by
Leo Szilard, who received a German patent for the device in
1929, but Szilard never attempted a practical realization of his
idea. Lawrence's own invention stemmed from his study of a paper by Rolf
Wideroe on resonant acceleration in linear structures using radio
frequency (RF) voltages. Although Lawrence could not understand
German, he was able to understand enough of the concept from the
drawings and equations to come to his own invention.  

In the Wideroe linear accelerator, a beam of ions was accelerated through a
series of small gaps between hollow metal tubes, called drift tubes,
connected in series to the RF-voltage generator. At any instant of time
successive gaps carry a voltage of opposite sign.  The voltage changes sign during the time it
takes the ion to traverse the tube, allowing
the ion to increase its energy.  Since the non-relativistic ions
increase their velocity in passing through the gaps, each successive
drift tube must be longer for the ion motion to remain in synchronism
with the RF-generator. Higher and higher energy meant the length 
of the accelerator increased non-linearly, at least initially.

To reach energies of 1~MeV or more, Lawrence's insight was that for
non-relativistic particles injected into a dipole magnetic field
perpendicular to the particle velocity, the revolution frequency is
independent of the particle energy. Higher energy particles travel on
larger orbits, maintaining synchronous (or isochronous) motion. If the
ion orbits are contained within two hollow ``D''-shaped electrodes
(``dees'') which are connected to an RF-voltage source, one can
accelerate ions to energies not possible with DC-voltage
structures. Beam is injected at the center of the 
device and spirals outward.

Lawrence's ideas were soon realized in practice for protons by his
student M. Stanley Livingston.  As has been the case since the first
accelerators, the Lawrence team pushed forward on two fronts: particle
energy and beam intensity.  The strong limitations on intensity were
imposed by losses of ions on the vertical surfaces of the cyclotron's
vacuum chamber.  This problem was mitigated when 
magnetic field shims were introduced to provide a radial component to the magnetic
field. The radial field increased with distance from the center of the cyclotron.
The result was vertical focusing that confined the beam to the median
horizontal plane.  With vertical focusing, many bunches of beam
particles, each with a different kinetic energy, could be
accelerated in the cyclotron. The cyclotron delivered a continuous
train of RF-bunches of ions.

Trying to increase the proton energy significantly beyond 10~MeV provided a
different difficulty.  The revolution frequency (cyclotron frequency)
began to decrease due to the relativistic increase in the proton mass.
Even a change as small as 1 to 
2\% is enough for the revolution frequency to be outside of the
frequency bandwidth of the RF-generator and therefore to lose
synchronism with the RF-voltage.  The solution to this limitation
seems simple: modulate the RF to lower values to maintain
synchronism with the highest-energy beam particles.  The frequency
modulated cyclotron (or synchrocyclotron) allowed the Lawrence team to
achieve energies as high as 340~MeV with their 184~inch 
cyclotron. 
At the University of Chicago, Enrico Fermi's team reached 450~MeV with
the slightly smaller 170~inch cyclotron. In the early 1950s this
machine produced copious pions via the $\Delta$(3,3)-resonance.  The age of accelerator-generated
neutrinos had begun.  It is noteworthy that Fermi and Szilard were
also responsible for the invention of another important source of man-made
neutrinos: the nuclear reactor.  

Even larger machines were built in Russia. Russia's effort culminated
in the giant 1~GeV proton synchrocyclotron at the Leningrad Institute for Nuclear
Physics in Gatchina. This machine was built around the world's largest
one-piece electromagnet with a pole diameter of 7~m and iron weight of
almost 8000~tonnes.\ Maintaining synchronism through frequency
modulation had come at a price.  Only one bunch, i.e., protons of only
a single energy, could be accelerated at any one time.  Therefore, the
Gatchina machine could only accelerate 0.2~$\mu$A, hundreds of times less
than classical fixed-frequency cyclotrons.

Generating continuous trains of ion bunches at energies exceeding $\sim$10
MeV required yet another invention that would allow particle
synchronism despite the acceleration being provided by fixed frequency
RF-power. The solution published by Llewellyn Thomas in 1938 showed
that conditions for resonance and focusing could be maintained if the
vertical magnetic field varied with polar angle. This variation
introduces an additional focusing effect and leads to scalloped rather
than simple spiral orbits.  If in addition to the azimuthal variation,
the vertical field increases in strength at larger radii, the
particle motion can be made isochronous – independent of energy – over
the full range of operation of the cyclotron. In isochronous
cyclotrons, the beam current can be increased to the mA range. Indeed,
1~mA  proton beams can be produced by commercially available machines
designed for radioisotope production today.

Donald Kerst introduced a further improvement of Thomas' scheme of azimuthal variation by radial
sectors. Kerst suggested using spiral sectors to increase axial focusing of the beam even more through the
application of the alternating gradient principle, which was by then
being designed into synchrotrons. Spiral sectors are now used in
almost all cyclotrons over $\sim$40 MeV, enormously increasing both the
energies and the intensities available and thereby providing a factor of 
$\sim$1000 more intense beams for $\pi$, $\mu$, n and neutrino production at low energies.

\begin{figure}[t]
\begin{center}
{\includegraphics[angle=-0, width=.95\linewidth]{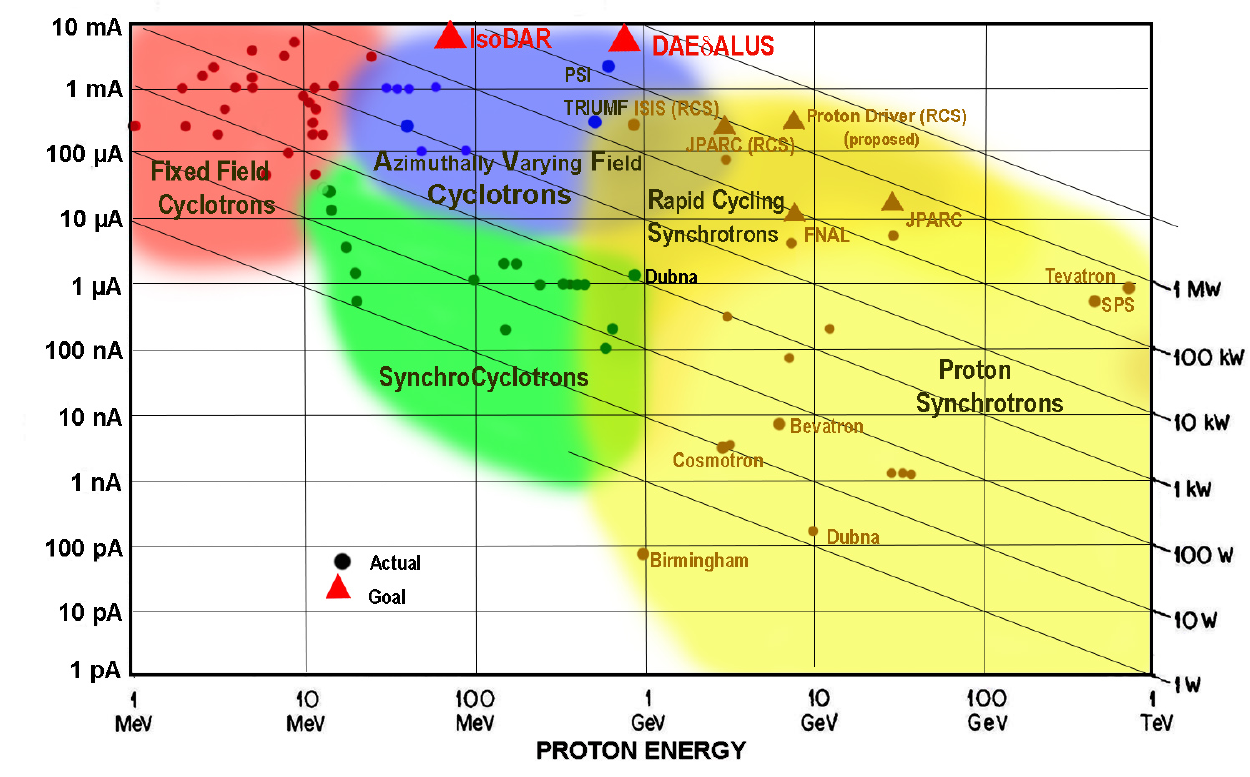}}
\end{center}
\vspace{-0.5cm}
\caption{Proton energy versus current for various existing machines.
  The type of accelerator is indicated. Various types of cyclotrons
  are noted, where  FF is the Fixed Field or Classical Cyclotron;
FM is the Frequency Modulation (Synchro-) Cyclotron; and 
AVF is the Azimuthal Varying Field Cyclotron. 
This plot is taken from Ref.~\cite{costeff}. 
\label{RCS} }
\end{figure}

The energy and intensity (or current) range provided by historical and
present-day cyclotrons, compared to other types of circular accelerators, is shown in Figure~\ref{RCS}.
Various types of cyclotrons are noted: FF is the Fixed Field or Classical Cyclotron;
FM is the Frequency Modulation (Synchro-) Cyclotron; and 
AVF is the Azimuthal Varying Field Cyclotron.  One can see that, at the
low energies needed for DAR beams,  cyclotrons are ideal drivers.
Linear accelerators are also an option, but require much higher power
and have much higher cost per unit energy than cyclotrons.

Research on two existing cyclotrons has provided important intellectual input for the 
800~MeV machine ultimately envisioned for the DAE$\delta$ALUS cyclotron program.   These 
are the Ring Cyclotron at the Paul Scherrer Institut (PSI) in Villigen,  Switzerland~\cite{psi-cycl-2010},
and the Superconducting Ring Cyclotron (SRC)~\cite{RIKEN} at RIKEN, Wako, Japan, which, although designed for high-energy
highly stripped heavy-ion beams, represents an engineering ``proof-of-principle'' design for a cyclotron magnet 
applicable for DAE$\delta$ALUS.    
With an energy of 590~MeV and beam current of 2.4~mA, the PSI Ring Cyclotron is currently the world's most powerful
accelerator in this energy range, delivering 1.4~MW of protons~\cite{psi-cycl-2010}, as seen in Figure~\ref{RCS}.\ The PSI complex routinely achieves 99.98\% extraction efficiency, and this sets the bar for future
accelerators such as those in the DAE$\delta$ALUS program.
The RIKEN SRC is the world's first
ring cyclotron that uses superconducting magnets, and has the strongest beam bending force among the cyclotrons.    
 The magnet design for the 800~MeV/n DAE$\delta$ALUS SRC is based on RIKEN.
RIKEN does not appear on Figure~\ref{RCS} because it is a heavy ion rather than a proton machine.  
As such, the current from the RIKEN machine is limited by the available shielding, and not by the machine design. 
RIKEN can boost the ion beam 
energy up to 440~MeV/nucleon for light ions and 350~MeV/nucleon for
very heavy ions, such as uranium nuclei, to produce intense radioactive
beams. The ring cyclotron consists of 
six major superconducting 
sector magnets with a maximum field of 3.8~T. The total stored energy
is 235~MJ, and its overall dimensions are 19~m diameter, 8~m height
and 8,300~tons. The magnet system assembly was completed in August
2005 and successfully reached the maximum field in November 2005. 
After magnetic field measurements for two months, the superconducting
magnets was installed and the first beam was extracted from the SRC in December 2006.

\subsection{Cyclotrons as pion/muon factories}
Cyclotrons have been used to produce pions and muons for many 
years; what is novel about DAE$\delta$ALUS is their application as 
drivers for DAR sources, in which the pions and muons come to rest 
and decay to neutrinos.    In fact, two out of three of the 
major ``meson factories''  commissioned in the 1970s  were 
cyclotron-based.  These were TRIUMF, in Vancouver, BC, Canada~\cite{Triumf}, and SIN~\cite{SIN} (now
PSI) in Villigen, Switzerland.    The
competing technology was LAMPF~\cite{LAMPFpi} (now LANSCE) at Los
Alamos, which was an 800~MeV linear accelerator.     These two 
cyclotron facilities remain at the forefront of precision pion and muon studies
to this day.  TRIUMF, a 500~MeV H$^-$ cyclotron, produces several hundred~$\mu$A.  
This program has expanded to also become a world-leading laboratory for
radioactive ion beams.  
PSI is currently the world's most powerful accelerator in this energy range with
590~MeV protons.  At 1.4~MW of beam power, this cyclotron is a shining
example of high-quality beams with extremely high extraction
efficiency (99.98\%) and low beam losses ($<$200 W per cyclotron
vault).    An example of the beautiful muon physics now being
published from PSI is the precision measurement of $G_F$ from the 
MuLan experiment \cite{MuLAN}.   However,  the PSI program with the
primary beam is now evolving toward being primarily directed at
production  of low-energy (meV) neutrons via the spallation process 
for neutron scattering and diffraction studies of materials.  These
cyclotrons, which have been a tremendous asset to the field, inform
the DAE$\delta$ALUS design below.

Although neither the TRIUMF nor PSI machines have been applied to the
neutrino field,
DAR experiments were conducted at Los Alamos with
the competing LAMPF beam
\cite{Chen, LSNDOsc}.  Also, it should be noted that other low-energy
synchrotrons have also hosted or are considering important neutrino
deca-at-rest experiments, namely the 700~MeV ISIS machine at the Rutherford
Appleton Lab in UK (KARMEN~\cite{KARMENOsc}) and the SNS at Oak Ridge
\cite{OscSNS}.  %A comparison of the beam energies and currents of these various accelerators is given in Fig.~\ref{fig:daed:cfg}.

\subsection{The DAE$\delta$ALUS cyclotrons}

\begin{figure}[t]
\begin{center}
\includegraphics[scale=.5]{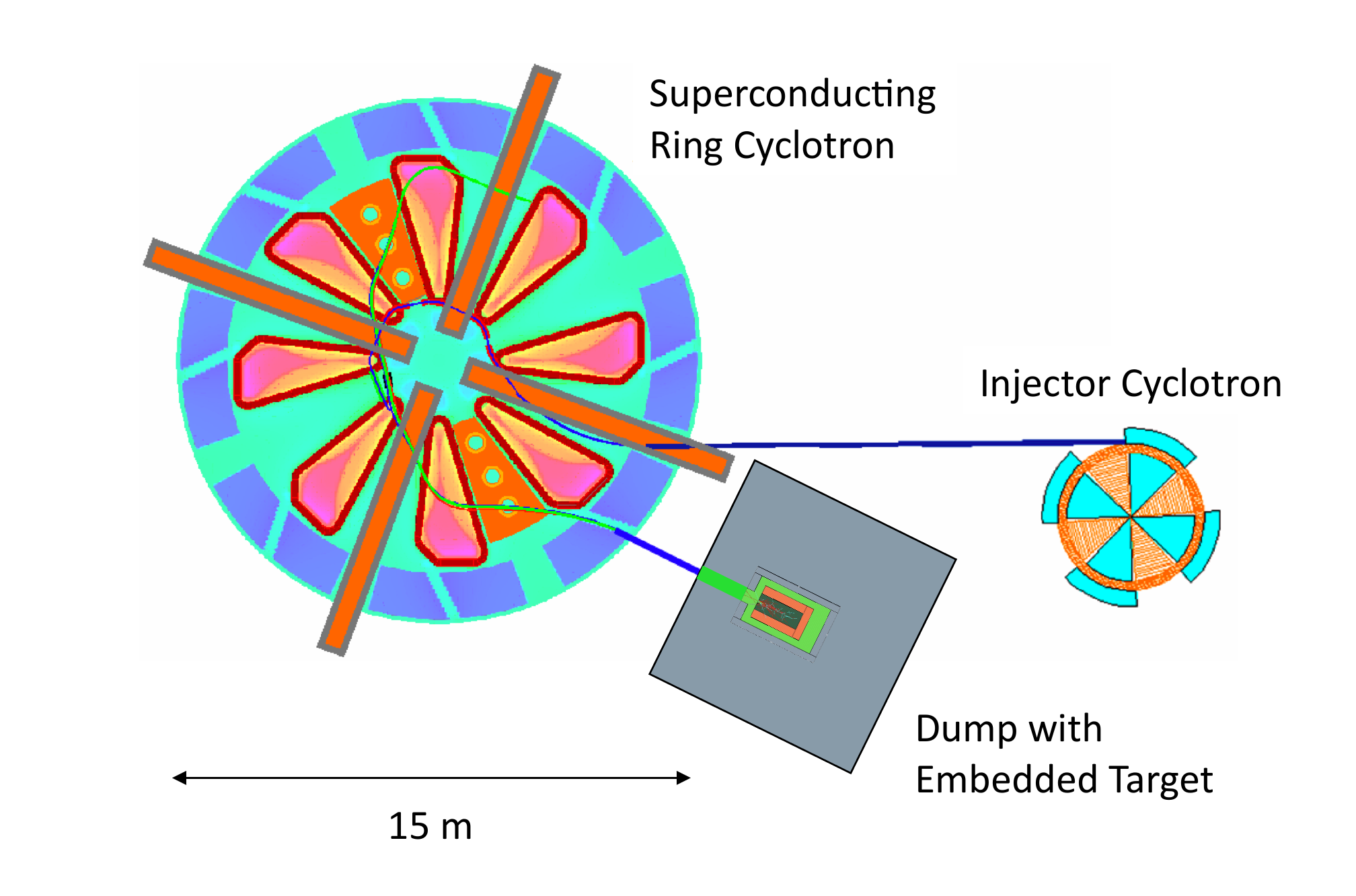}
\end{center}
\vspace{-.5cm}
\caption{A DAE$\delta$ALUS module.} 
\label{fig:daed:cfg}
\end{figure}

The DAEdALUS cyclotron system accelerates ions through a series of two
cyclotrons.    The full system is called an ``accelerating module.''
Figure~\ref{fig:daed:cfg} shows the schematic layout of one of the
\htp\ accelerating modules.  The DAE$\delta$AUS injector cyclotron
(DIC) captures
up to 5~mA (electrical) of \htp\ and accelerates the beam to about
60~MeV/n. This beam is extracted electrostatically.    This beam can
then be used for a stand-alone program (IsoDAR) or for 
injection into the DSRC for further acceleration (DAE$\delta$ALUS). This second machine consists of 8~wedge-shaped superconducting magnets and 6~RF cavities (4~of
the PSI single-gap type, 2~double-gap). The stripper foil is located at the outer radius, at the trailing edge of one of the sector magnets, 
and the extraction channel comes out roughly along one of the valleys about 270$^o$ away. Figure \ref{fig:daed:cfg} also shows schematically one of the beam dumps, a graphite block with a hole shaped to correspond to the
beam profile so the energy is uniformly distributed over a wide area. The graphite is surrounded by a copper, water-cooled jacket and is expected to dissipate 6~MW of beam power.

A key aspect of the design is acceleration of \htp.
This novel choice of ion was selected by L. Calabretta~\cite{Lucianofirstpaper} in response to a
suggestion by Carlo Rubbia in the 1990s to use high-current, $\sim 1$
GeV cyclotrons for driving thorium reactors~\cite{Rubbia}.    Since
most cyclotrons accelerate protons or \hmi, it is worth examining the
pros and cons of this choice.

A drawback of \htp\ is that the higher rigidity of this ion ($q/A$=0.5) compared to bare protons 
or \hmi (both with $|q|/A$=1) 
requires a cyclotron of relatively large radius. However, the size of
the machine is practical given the higher fields available from superconducting magnets. 
In fact, the RIKEN SRC is close to the field and size specifications required.

By choosing the \htp\ ion,  one can use a stripping foil to cleanly extract the beam from the cyclotron.  
Although stripping extraction is not available to proton accelerators, it has been used extremely effectively 
in lower-energy cyclotrons that accelerate \hmi\ beams.  The value of stripping extraction is that turn separation
 is no longer an issue. All ions will pass through the stripper foil,
even if turn separation is not clean. Ions will be stripped from
several turns, 
probably not more than two or three, and the protons will carry the
energy associated with their turn number. 
The extraction channel, which will pass through 
the central region of the DSRC (as the protons are bent in rather than
out as is the case with \hmi), 
will need adequate momentum acceptance to transmit all the protons
from the turn where they are stripped.

It is the lower binding energy (0.7~eV) of the \hmi\ ion that renders
it unusable in a high-energy machine like the DSRC.
This makes \hmi\ susceptible to Lorentz stripping in fields as low as 2~T and
energies below about 70~MeV. The higher binding energy of the \htp\ ion (2.7~eV, at least in its ground state) renders it more
stable, and able to survive to 800~MeV in the highest 6~T fields anticipated in the DSRC.

Lastly, a great advantage of \htp\ is reduction of space-charge
effects. The space charge of the particle beam produces a repulsive
force inside the beam, 
which generates detuning effects. A measure of the strength of this effect, the ``generalized perveance,'' is defined by
\begin{equation} \label{eq:per}
K = \frac{qI}{2\pi\epsilon_0 m c^3 \gamma^3 \beta^3},
\end{equation}
where $q, I, m, c, \gamma$, and $\beta$ are the charge, current, rest mass, speed of light, and the relativistic
parameters of the particle beam, respectively~\cite{reiser2008theory}. The higher the value of $K$, the stronger the space-charge detuning effects.

According to Eq.~\ref{eq:per}, the space-charge effects for the 5~mA
of \htp\ beam in the DSRC are equivalent to a 2.5~mA proton beam with
the same  $\gamma$. Consequently, they are similar to the space-charge
effects present in the 2.4~mA proton beam currently being accelerated
at PSI. Another degree of freedom to reduce space-charge effects are
the choices injection energy and acceleration voltage. 
Given this premise, we have carried out precise beam dynamics studies,
including the 3D space-charge effects (excluding the central region of the
DIC), with self-consistent 3D models implemented in the code Object
Oriented Parallel Accelerator Library (\opal)~\cite{opal}. The beam
dynamics model is described in detail in~\cite{PhysRevSTAB.13.064201,
  PhysRevSTAB.14.054402}. For the DSRC, we have implemented a simple
stripper model into \opal\ in order to study the complex extraction
trajectories of the stripped protons.    

DAE$\delta$ALUS R\&D related to \htp\
acceleration has begun at a test stand at Best Cyclotrons, Inc., in
Vancouver, BC, Canada.    These studies employ the  VIS, or Versatile
Ion Source, a non-resonant ECR (Electron Cyclotron Resonance) source
\cite{Maimone:2008zz} built at the Laboratori Nazionali del Sud (LNS)
in Catania, Italy, which has been shipped to Vancouver for these tests.
As with any ion source,  protons, H$_2^+$ and even H$_3^+$ will be
emitted.   We have begun studies of emittance, inflection into the
cyclotron, capture, and acceleration of H$_2^+$. 

In summary, the DAE$\delta$ALUS program has developed a plan for
the production of two high power cyclotrons,  one producing beam at 60~MeV/n and the other at 800~MeV/n.  
The former provides the injector to the latter, which then can produce pion/muon DAR neutrino fluxes.  
As discussed below, the injector can also be used by itself to produce isotope DAR beams.

\subsection{Application of the DIC Cyclotron to Isotope Production}

The remainder of this paper will describe the value of the DAE$\delta$ALUS
cyclotrons for basic research in neutrino physics.    However, it is
worth pausing to note that these cyclotrons have
practical applications.  This has attracted industry
involvement in DAE$\delta$ALUS development.
Examples range from medical applications to accelerator driven systems
for thorium reactors.   We concentrate on the former here, and note that
Ref.~\cite{Alonso:2012ub} provides a more extensive description of isotope production
for medicine using the DIC cyclotron.

Even from Ponetcorvo's days, isotopes produced by cyclotrons were being
investigated for suitability in medical diagnostics and therapeutics.  
The first direct therapeutic use of beams occurred in 1937 with trials
using neutrons produced from the 60~MeV Crocker Cyclotron at Berkeley.  
Incidentally, this machine is still in use, primarily for proton
treatments of eye tumors at UC Davis.  
Higher energy cyclotrons were built in the 1940's, the 800~MeV
184~inch Synchrocyclotron at Berkeley, 
and two similar-sized machines at Dubna and St. Petersburg.  
Medical treatments played prominent roles in all three, first with
stereotactic microsurgery with very fine beams for pituitary ablation,
then when diagnostic tools such as CT scanners became available that could
carefully measure the density of material upstream of a tumor to
accurately predict the stopping point of the beam, Bragg-peak therapy 
with alpha-particle beams was instituted at the 184~inch
Synchrocyclotron.
 
Cyclotrons are now used extensively for therapy with proton beams,
with several commercial companies marketing highly effective
cyclotron-based systems for proton therapy with beam energies of
around 250~MeV \cite{IBA}.  As beam currents needed for radiotherapy
are only in the nA range, even with the losses inherent in degrading
the fixed-energy beams down to match the required range in the
patient, adequate beam brightness can be achieved with beams 
of no more than a $\mu$A of protons extracted.  

Meanwhile, radioactive isotopes produced with cyclotrons of energies
of 30~MeV or less have become widely used in medical diagnostics and
therapy, 
with an ever-increasing demand as techniques are refined and results improved.
Beam currents in the range of 750~$\mu$A to 2~mA are now
being extracted from commercial and research isotope-producing
cyclotrons; the limiting factor is often heat-dissipation in the
complex targets that are needed for effective isotope production.
Increased production capacity is being obtained by multiple extraction
ports enabled by acceleration of H$^-$ beams that can be extracted by
stripping.  Sharing the total beam power between two 
target stations enables greater production capacity.

\begin{table}[t]
\centering
{
\begin{tabular}{|l|c|c|}
\hline
Isotope & half-life & Use \\ \hline
$^{52}$Fe & 8.3 h &  The parent of the PET isotope $^{52}$Mn  \\
   & & and iron tracer
  for red-blood-cell formation and brain uptake studies.\\  \hline
$^{122}$Xe & 20.1 h &  The parent of  PET isotope $^{122}$I used to study
brain blood-flow. \\ \hline
$^{28}$Mg  & 21 h & A tracer that can be used for bone studies,
analogous to calcium \\ \hline
$^{128}$Ba  & 2.43 d & The parent of positron emitter $^{128}$Cs. \\
  & & As a
  potassium analog, this is used for heart and blood-flow imaging. \\ \hline
$^{97}$Ru & 2.79 d & A $\gamma$-emitter used for spinal fluid and liver
studies. \\ \hline
$^{117m}$Sn & 13.6 d & A $\gamma$-emitter potentially useful for bone
studies. \\ \hline
$^{82}$Sr & 25.4 d &  The parent of positron emitter $^{82}$Rb, a
  potassium analogue   \\ 
& &  This isotope is also directly used as a PET
  isotope for heart imaging. \\ 
\hline
\end{tabular}}
\caption{  Medical isotopes relevant at IsoDAR
  energies, reprinted from Ref.~\cite{costeff}. }
\label{tab:med}
\end{table}

The DAE$\delta$ALUS injector cyclotron, used for IsoDAR, 
will become a
powerful tool for isotope production along two different directions.
As a source of 60~MeV protons at 600~kW, beam powers are substantially
higher than existing isotope machines.  This could enable either
significantly greater yield on a single target, should the technology
be developed to use all this beam power on a target or by sharing the
beam between many targets to increase the versatility of the isotope
factory.  As H$_2^+$ ions are extracted from the cyclotron via a
conventional septum, a narrow stripper can be placed over a portion of
the beam to convert ions passing through the stripper into protons
that can then be cleanly separated from the body of the beam and
transferred to a production target.  The remaining beam is transported
to further stripping stations, each peeling of a 
small portion of the beam to deliver to a different target.  
In this way the power limits on any given target will not be exceeded,
and high efficiency for use of the whole beam 
maintained.  Examples of the isotopes which can be produced, 
and their applications, are shown in Table~\ref{tab:med}. 

A second isotope application of the H$_2^+$ cyclotron is that ions of
the same charge-to-mass ratios can also be accelerated.  Specifically,
He$^{++}$ (alpha-particle) beams can be accelerated at currents
limited only by the availability of such He$^{++}$ ion sources.  There
are many isotopes that have tremendous application potential and are
limited today only by the very restricted availability of suitable
high-current alpha beams.  In fact, the first prototype cyclotron to
be built for testing injection of the high-current H$_2^+$ beams, to
be built at the LNS in Catania, Italy, is being designed to be used
directly following the H$_2^+$ injection tests as a 
dedicated alpha-particle cyclotron for producing radiotheraputic
isotopes. 
One example will be $^{211}$At, which is in short supply for even long term clinical studies

The DAE$\delta$ALUS Superconducting Ring Cyclotron, in extending the
performance of today's record-holding 
PSI by increasing energy from 590 to 800~MeV and a factor of five in
current, becomes a member of the GeV - 10~MW - class of machines. Many
such machines have been designed and proposed but cost has been an
impediment to their realization.  To date, only one such project has
progressed to the advanced R\&D and construction phase:
MYRRHA~\cite{myrrha} to be sited in Mol, Belgium.  These projects all
fall within the ADS (Accelerator-Driven Systems) category, 
such as nuclear waste transmutation, driving of sub-critical
thorium-based reactors,
 tritium production, and others.  

Along with the physics possibilities previously described, the
DAE$\delta$ALUS cyclotrons 
provide new opportunities in this field by offering beams at a substantially reduced cost over the linear accelerators which until now have been viewed 
as the only viable technology to reach these levels of beam power in
the GeV energy range.  
With successful development of these cyclotrons, a substantial growth in the ADS field can be anticipated, with the cost hurdle having been surpassed.

\section{IsoDAR}
IsoDAR is the first proposed step of DAE$\delta$ALUS. IsoDAR represents both a novel concept of application to neutrino physics measurements and a demonstration of the 60~MeV/n injector cyclotron relevant for the larger program. 

The baseline cyclotron design for IsoDAR is a 5~mA H$_{2}^{+}$ machine that will
accelerate beam to 60~MeV per nucleon. Beam would be injected at 70~keV (35~keV/amu) via a spiral inflector. 
An appropriate ion source already exists and its
testing will be completed in Summer~2013. 
The current plan for IsoDAR is to locate a cyclotron accelerator underground in an experimental hall close to the KamLAND detector, in Kamioka, Japan. 
This is
a continuous-wave source with a 90\% duty cycle to allow for
machine maintenance. The resulting beam will be transported a short
distance up the drift at KamLAND to a target located in a room. The end of the beam dump
 is assumed to be 16~m from the detector center. 

We continue to optimize the target for the production of $^{8}$Li, a
$\beta$-emitter and the source of the $\overline{\nu}_e$ for the
measurement.  The baseline design for the target is a cylinder of
$^{9}$Be, 20~cm long and 20~cm in diameter. This cylinder is surrounded
by an additional 5~cm of D$_{2}$O which works to both moderate neutrons and to provide target cooling. The  D$_{2}$O is then
surrounded by a cylindrical sleeve of 99.99\% pure $^{7}$Li, 150~cm
long and 200~cm in outer diameter. Some $^{8}$Li is produced directly
in the $^{9}$Be target but the majority of the $^{8}$Li is produced by
the many neutrons made in the $^{9}$Be target capturing on $^{7}$Li. The
isotopic purity of the $^{7}$Li sleeve is needed to avoid production
of tritium by neutrons on $^{6}$Li. Further, this production cross section is several orders of magnitude larger than neutron capture on $^{7}$Li and
therefore reduces the production of $^{8}$Li severely. The needed
level of $^{7}$Li purity for the sleeve is readily available from a
number of sources as it is commonly used in the nuclear industry.
A nominal running period of five years with a 90\% duty cycle produces
1.29$\times$10$^{23}$ antineutrinos from the decay of $^{8}$Li.

When paired with a liquid scintillator detector, this isotope DAR flux opens a number of opportunities for precision neutrino measurements.  This paper presents two examples.  The first is a high sensitivity
sterile neutrino search.   The second is a search for new physics in the neutrino sector from neutrino-electron scattering.  Both cases describe pairing with KamLAND, to provide specific
information on rates and backgrounds. An example involving the detection of coherent neutrino scattering is also provided although such a measurement would require a dark-matter-style detector sensitive to keV-scale excitations.

\subsection{Sterile neutrino searches}
Searches for light sterile neutrinos with mass $\sim$1~eV are motivated by observed
anomalies in several experiments. Intriguingly, these results come from a wide range of experiments covering neutrinos, anti-neutrinos, different flavors, and different energies. Short baseline accelerator
neutrino oscillation experiments~\cite{LSNDOsc,Aguilar-Arevalo:2013pmq}, short
baseline reactor experiments~\cite{Mueller,Lassere}, and even the radioactive source experiments, which were originally intended as calibrations for the chemical solar neutrino experiments~\cite{SAGE, GALLEX}, have all 
observed anomalies that can be interpreted as due to one or more sterile neutrinos. To understand these anomalies in terms of the $\nu$SM for neutrino oscillations, one or more sterile neutrinos are added to the
oscillation probability calculation~\cite{Sorel:2003hf}. These extended models are
referred to as ``3+1", ``3+2", or ``3+3" neutrino models depending on the number of additional sterile 
neutrinos. Global fits to these data indicate that there are regions in this extended parameter space where
the anomalies can be reconciled with each other as well as negative results from other experiments in and near the relevant parameter space~\cite{sterile2013}. The global fits tend to prefer models with two or three sterile neutrinos.

The exciting diversity of experiments showing these anomalous results has motivated a number of proposals to address them. Suggestions range from repeating the source experiments, to specially designed reactor antineutrino experiments, to accelerator-based ones. Many of these proposals, however, do not have sufficient
sensitivity to make a definitive $>5\sigma$ statement about the
existence of sterile neutrinos in all of the relevant parameter space. The experiments that are designed to make a definitive measurement are based on pion or isotope DAR sources. Notably, the full DAE$\delta$ALUS complex
could be used to generate a pion DAR beam for such a
measurement.  However, the IsoDAR concept calls for simply the DAE$\delta$ALUS injector cyclotron to be used to generate an isotope DAR source. Such a complex situated next to a kiloton-scale scintillator detector such as
KamLAND would enable a definitive search for sterile
neutrinos by observing a deficit of antineutrinos as a function of the distance $L$ and antineutrino energy $E$ across the detector--the definitive
signature of neutrino oscillation. This is the concept behind the
IsoDAR proposal~\cite{Bungau:2012ea}.

The proposed IsoDAR target is to be placed adjacent to the KamLAND detector.
The antineutrinos propagate 9.5~m through a combination of rock, outer muon
veto, and buffer liquid to the active scintillator volume of KamLAND. The scintillator is contained 
in a nylon balloon 6.5~m in radius bringing the total distance from target to detector center to
16~m. The antineutrinos are then detected via the IBD interaction. This interaction has a well known cross section with an
uncertainty of 0.2\%~\cite{Vogel:1999zy}, and creates a distinctive coincidence signal between a prompt positron signal, $E_{e^+}=E_{\bar{\nu}_{e}} - 0.78$~MeV, and a delayed neutron capture giving a 2.2~MeV gamma ray within $\sim$200~$\mu$s. 

KamLAND was designed to efficiently detect IBD. A standard analysis has a 92\%
efficiency for identifying IBD events~\cite{Gando:2010aa}. In IsoDAR's
nominal 5 year run, $8.2\times10^{5}$ IBD events are expected. The
largest background comes from the 100~reactor antineutrino IBD events
detected by KamLAND per year~\cite{Abe:2008aa}. The reactor
antineutrino rate is dependent on the operation of the nuclear
reactors in Japan which has been significantly lower in 2012 and
2013~\cite{Gando:2013nba}. The sterile neutrino analysis uses an
energy threshold of 3~MeV. Due to the effective background rejection
efficiency provided by the IBD delayed coincidence signal, this
threshold enables use of the full KamLAND fiducial volume, R$<$6.5~m
and 897~tons, with negligible backgrounds from sources other than from the
aforementioned reactor antineutrinos.

\begin{table}
\begin{center}
\begin{tabular}{ l l}
\hline
\hline
Parameter & Value \\
\hline
Run period & 5 years (4.5 years live time) \\
$\bar{\nu}_{e}$ flux & $1.29\times10^{23}$ $\bar{\nu}_{e}$  \\
Fiducial mass & 897 tons \\
Target face to detector center & 16 m \\
Detection efficiency & 92\% \\
Vertex resolutions & 12 cm/$\sqrt{E(MeV)}$ \\
Energy resolutions & 6.4\%/$\sqrt{E(MeV)}$ \\
Prompt energy threshold & 3 MeV \\
IBD event total & $8.2\times10^5$ \\
\hline
\hline
\end{tabular}
\caption{\label{parSterile}The KamLAND detector parameters used in calculating the sterile neutrino search sensitivity.}
\end{center}
\end{table}

\begin{figure}[tb]
\begin{center}
\includegraphics[scale=.7]{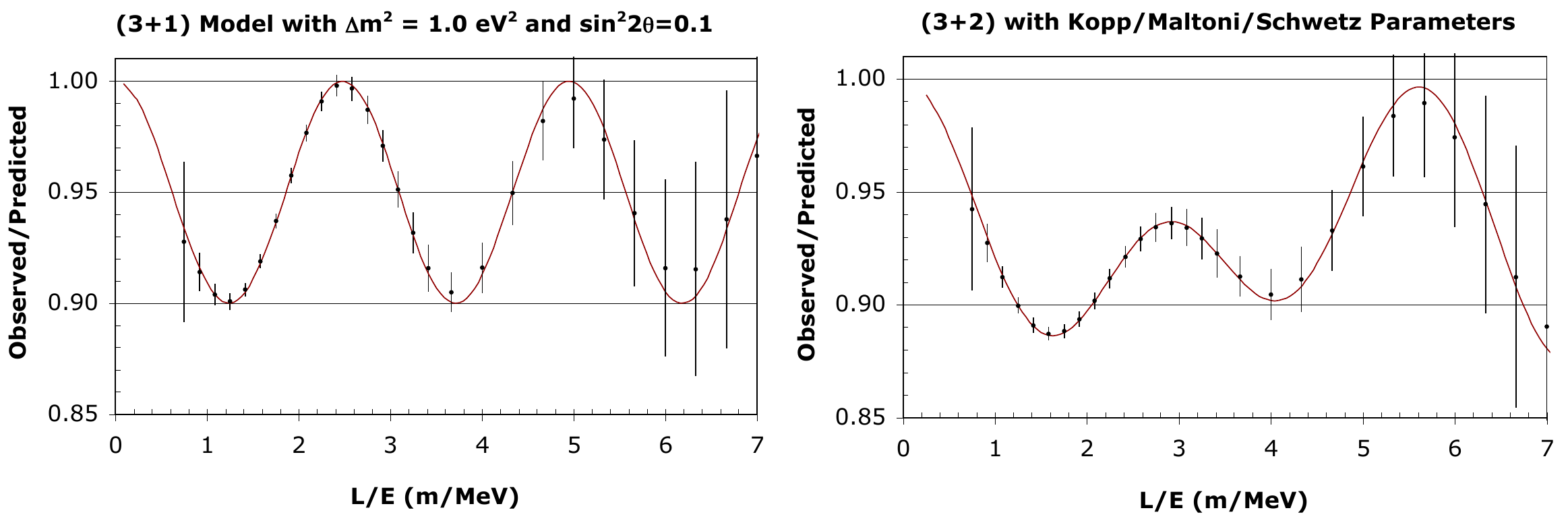}
\end{center}
\vspace{-.5cm}
\caption{Example data sets for 5 years of running for 3+1 (left) and 3+2
  (right) oscillation scenarios. 
\label{waves}}
\end{figure}

The sterile neutrino analysis makes use of neutrino oscillations's $L/E$ signature. Therefore, the energy and vertex 
resolutions are essential in determining sensitivity. The KamLAND detector has a vertex reconstruction resolution 
of $12{~\rm cm}/\sqrt{E(\rm MeV)}$ and an energy resolution of  $6.4\%/\sqrt{E(\rm MeV)}$~\cite{Gando:2010aa}.  
Example data sets for reasonable 3+1 and 3+2 sterile are shown in Figure~\ref{waves} for the nominal detector
 parameters, summarized in Table~\ref{parSterile}. In most currently favored oscillation scenarios, the $L/E$ signal is observable. Furthermore, separation of the various 3+N models may be possible as exemplified by Figure~\ref{waves} (right).

\begin{figure}
\begin{center}
\includegraphics[width=0.6\columnwidth]{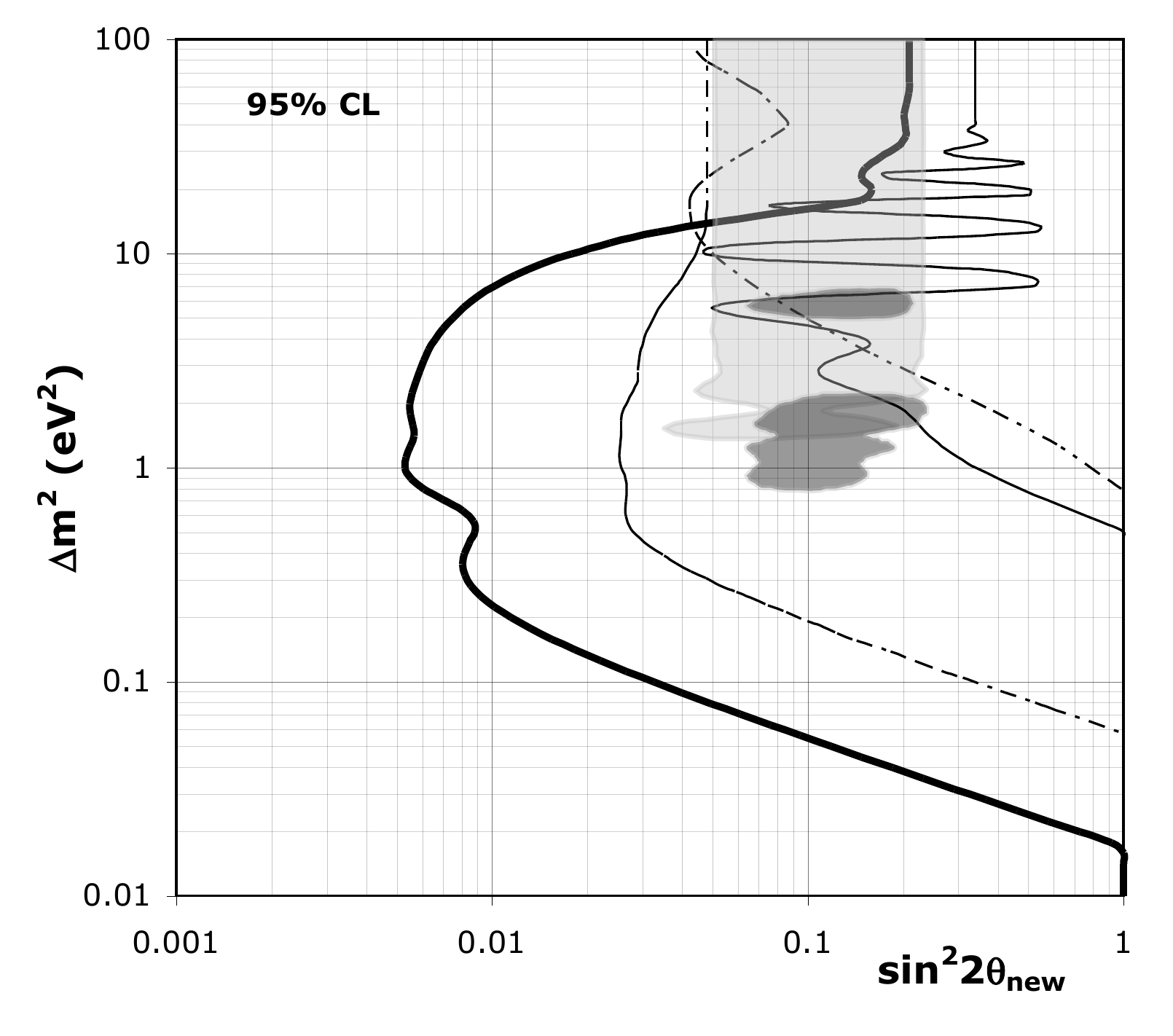} 
\vspace{-.2cm}
\caption{Sensitivity of IsoDAR in a nominal 5 year run, in comparison
  with other experiments\label{sensitivySterile}.}
\end{center}
\end{figure}

To understand the sensitivity relative to other proposals,
the IsoDAR 95\% CL is compared to other electron antineutrino disappearance experiments in the two
neutrino oscillation parameter space in Figure~\ref{sensitivySterile}. In just five years of running, IsoDAR rules out the entire global 3+1 allowed region, $\sin^2{2\theta_{new}}=0.067$ and $\Delta
m^{2}=1$~eV$^{2}$ at 20$\sigma$. This is the most definitive
measurement among the proposals in the most probable parameter space
of $\Delta m^{2}$ between $1-10$~eV$^{2}$.

\subsection{Precision electroweak tests of the Standard Model}
In addition to the $8.2\times10^5$ IBD interactions, the IsoDAR neutrino source~\cite{PRL}, when combined with the KamLAND detector~\cite{KL}, can collect the largest sample of low-energy $\bar \nu_e$-electron (ES) scatters that has been observed to date. More than 7200 ES events will be collected above a 3~MeV visible energy threshold over a 5 year run, and both the total rate and the visible energy can be measured. This can be compared to the samples from  the Irvine experiment (458 events from 1.5 to 3~MeV~\cite{Irvine:1976}); TEXONO (414~events from 3 to 8~MeV~\cite{TEXONO:2012}); Rovno (41~events from 0.6 to 2~MeV~\cite{ROVNO:1993}); and MUNU (68 events from 0.7 to 2~MeV~\cite{MUNU:2005}).  

In the Standard Model, the ES differential cross section is given by

\begin{equation}
\frac{d\sigma}{dT} = \frac{2 G_F^2 m_e}{\pi}
\left[ 
g^2_R + g^2_L\left(1 - \frac{T}{E_\nu}\right)^2 - g_R g_L \frac{m_e T}{E^2_\nu}
\right],
\label{glgrxs}
\end{equation}

\noindent where $T\in\left[0,\frac{2E_{\nu}^2}{m_e+2E_{\nu}}\right]$ is electron recoil energy, $E_\nu$ is the energy of the incoming $\bar\nu_e$, and the weak coupling constants $g_R$ and $g_L$ are given at tree level by $g_R = \sin^2\theta_W$ and $g_L=\frac{1}{2}+\sin^2\theta_W$.  Eq.~\ref{glgrxs} can also be expressed in terms of the vector and axial weak coupling constants, $g_V$ and $g_A$, using the relations $g_R=\frac{1}{2}(g_V-g_A)$ and $g_L=\frac{1}{2}(g_V-g_A)$.

The ES cross section can be therefore be used as a probe of the weak couplings, $g_V$ and $g_A$, as well as $\sin^2\theta_W$, a fundamental parameter of the Standard Model as described in Ref.~\cite{sinsqthw}.  Although $\sin^2\theta_W$ has been determined to high precision~\cite{ewwg}, there is a longstanding discrepancy~\cite{PDG} between the value obtained by $e^+e^-$ collider experiments and the value obtained by NuTeV, a precision neutrino-quark scattering experiment~\cite{NUTEV:2002}.  Despite having lower statistics than the NuTeV, IsoDAR would measure $\sin^2\theta_W$ using the purely leptonic ES interaction, which does not involve any nuclear dependence.  This could therefore shed some light on the value of $\sin^2\theta_W$ measured by neutrino scattering experiments.

The ES cross section is also sensitive to new physics in the neutrino sector arising from nonstandard interactions (NSIs), which are included in the theory via dimension six, four-fermion effective operators.  NSIs give rise to weak coupling corrections and modify the Standard Model ES cross section given in Eq.~\ref{glgrxs} to
\begin{equation}\label{epsilonxsec}
\frac{d\sigma}{dT}= \frac{2 G_F^2 m_e}{\pi} [ (\tilde g_R^2+\sum_{\alpha \neq e}|\epsilon_{\alpha e}^{e R}|^2)+(\tilde g_L^2+\sum_{\alpha \neq e}|\epsilon_{\alpha e}^{e L}|^2)\left(1-{T \over E_{\nu}}\right)^2-(\tilde g_R \tilde g_L+ \sum_{\alpha \neq e}|\epsilon_{\alpha e}^{e R}||\epsilon_{\alpha e}^{e L}|)m_e {T \over E^2_{\nu}}],
\end{equation}
\noindent where $\tilde g_R= g_R+\epsilon_{e e}^{e R}$ and $\tilde g_L=g_L+\epsilon_{e e}^{e L}$. The NSI parameters $\epsilon_{e\mu}^{e LR}$ and $\epsilon_{e\tau}^{e LR}$ are associated with flavor-changing-neutral currents, whereas $\epsilon_{ee}^{e LR}$ are called non-universal parameters.  We can estimate IsoDAR's sensitivity to these parameters by fitting Eq.~\ref{epsilonxsec} to the measured ES cross section, assuming the Standard Model value for $\sin^2\theta_W$.  In general, lepton flavor violating processes are tightly constrained so we focus only on IsoDAR's sensitivity to the two non-universal parameters $\epsilon_{ee}^{e LR}$.

The ES interaction used for these electroweak tests of the Standard Model is very different than the IBD interaction used for the sterile neutrino search. The IBD signal consists of a delayed coincidence of a positron and a 2.2 MeV neutron capture $\gamma$, whereas the ES signal consists of isolated events in the detector.  Another difference is that at IsoDAR energies, the IBD cross section is several orders of magnitude larger than the ES cross section.  In fact, if just 1\% of IBD events are mis-identified as ES events, they will be the signal largest background.  On the other hand, as was suggested in Ref.~\cite{Conrad:2004gw}, the IBD signal can also be used to reduce the normalization uncertainty on the ES signal to about 0.7\% .  A final difference is that the incoming $\bar\nu_e$ energy for IBD interactions in KamLAND can be inferred from the visible energy on an event-by-event basis, while the incoming $\bar\nu_e$ energy for ES interactions in KamLAND cannot.  Therefore, the differential ES cross section is measured in visible energy bins corresponding to the kinetic energy of the recoil electron, and the dependence on the incoming $\bar\nu_e$ energy is integrated out according to the IsoDAR flux.

\begin{table}[tb]
\begin{center}
\begin{tabular}{c|c}
\hline
           &     Events \\
\hline
   Elastic scattering (ES) &       2583.5 \\

IBD Mis-ID Background &        705.3 \\

Non-beam Backkground &       2870.0 \\
\hline
Total   &  6158.8  \\
\hline
\end{tabular}     
\end{center}
\caption{ {Total signal and background events in KamLAND with $E_{\mathrm{vis}}$ between 3--12 MeV given the IsoDAR assumptions in Table~\ref{parSterile} and the selection cuts outlined in text.} \label{events}}
\end{table}

The backgrounds to the ES signal can be grouped into beam-related backgrounds, which are dominated by mis-identified IBD events, and non-beam backgrounds, arising from solar neutrino interactions, muon spallation, and environmental sources. We adopt a strategy similar to the one outlined in Ref.~\cite{kamBoron} to reduce the non-beam backgrounds.  First, a cosmic muon veto is applied to reduce the background due to radioactive light isotopes produced in muon spallation inside the detector. This reduces the live time by 62.4\%.  Next, a visible energy threshold of 3 MeV is employed to reduce the background from environmental source which pile up at low energies.  Finally, a fiducial radius of 5 m is used to reduce the background from external gamma rays emanating from the rock or stainless steel surrounding the detector.  To reduce the beam-related backgrounds, an IBD veto is employed to reject any ES candidate that is within 2 ms of a subsequent event with visible energy $>1.8~\mathrm{MeV}$ in a 6 m fiducial radius.  The IBD veto is estimated to have an efficiency of $99.75\% \pm 0.02 \%$, where the uncertainty is assumed to come from the statistical uncertainty on measuring the IBD selection efficiency with 50,000 AmBe calibration source events.

\begin{table}[tb]
  \begin{center}
\begin{tabular}{r|cccc}
\hline
           &      Background factor & $\delta \sin^2\theta_W$ &$ \frac{\delta \sin^2\theta_W}{\sin^2\theta_W} $
           &  $\delta \sin^2\theta_W^\text{stat-only}$  \\
\hline
Rate +Shape &        1.0 &     0.0076 &      3.2\% &     0.0057 \\

Shape Only &        1.0 &     0.0543 &     22.8\% &     0.0395 \\

 Rate Only &        1.0 &     0.0077 &      3.2\% &     0.0058 \\

Rate +Shape &        0.5 &     0.0059 &      2.5\% &     0.0048 \\

Rate +Shape &        0.0 &     0.0040 &      1.7\% &     0.0037 \\
\hline
\end{tabular}      
\end{center}
\caption{ {Estimated $\sin^2\theta_W$ measurement sensitivity for various types of fits to the E$_{\mathrm{vis}}$ distribution.  The second column indicates the background reduction factor.} \label{results}}
\end{table}

\begin{figure}[tb ]
	\begin{center}		
		\includegraphics[width=0.49\textwidth]{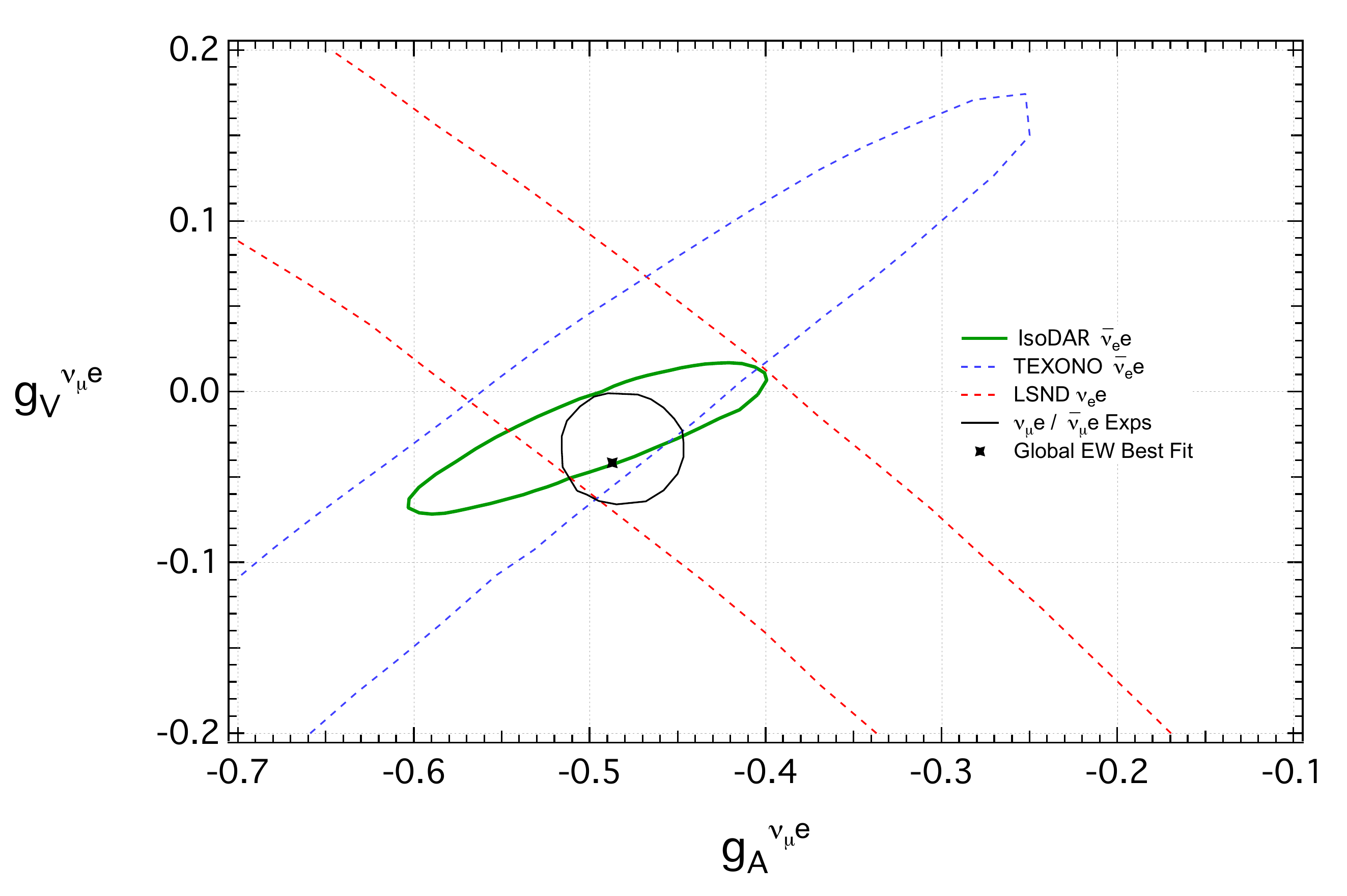}
	\end{center}
	\caption{\label{gV_gA} IsoDAR's sensitivity to $g_V$ and $g_A$ along with allowed regions from other neutrino scattering experiments and the electroweak global best fit point taken from Ref.~\cite{PDG}. The IsoDAR, LSND, and TEXONO contours are all at $1\sigma$ and are all plotted in terms of $g_{V,A}^{\nu_\mu e}= g_{V,A}^{\nu_e e}-1$ to compare with $\nu_\mu$ scattering data. The $\nu_\mu e/\bar{\nu}_\mu e$ contour is at 90\% C.L.}
	\end{figure}
\begin{figure}[tbh ]
	\begin{center}		
		\includegraphics[width=0.49\textwidth]{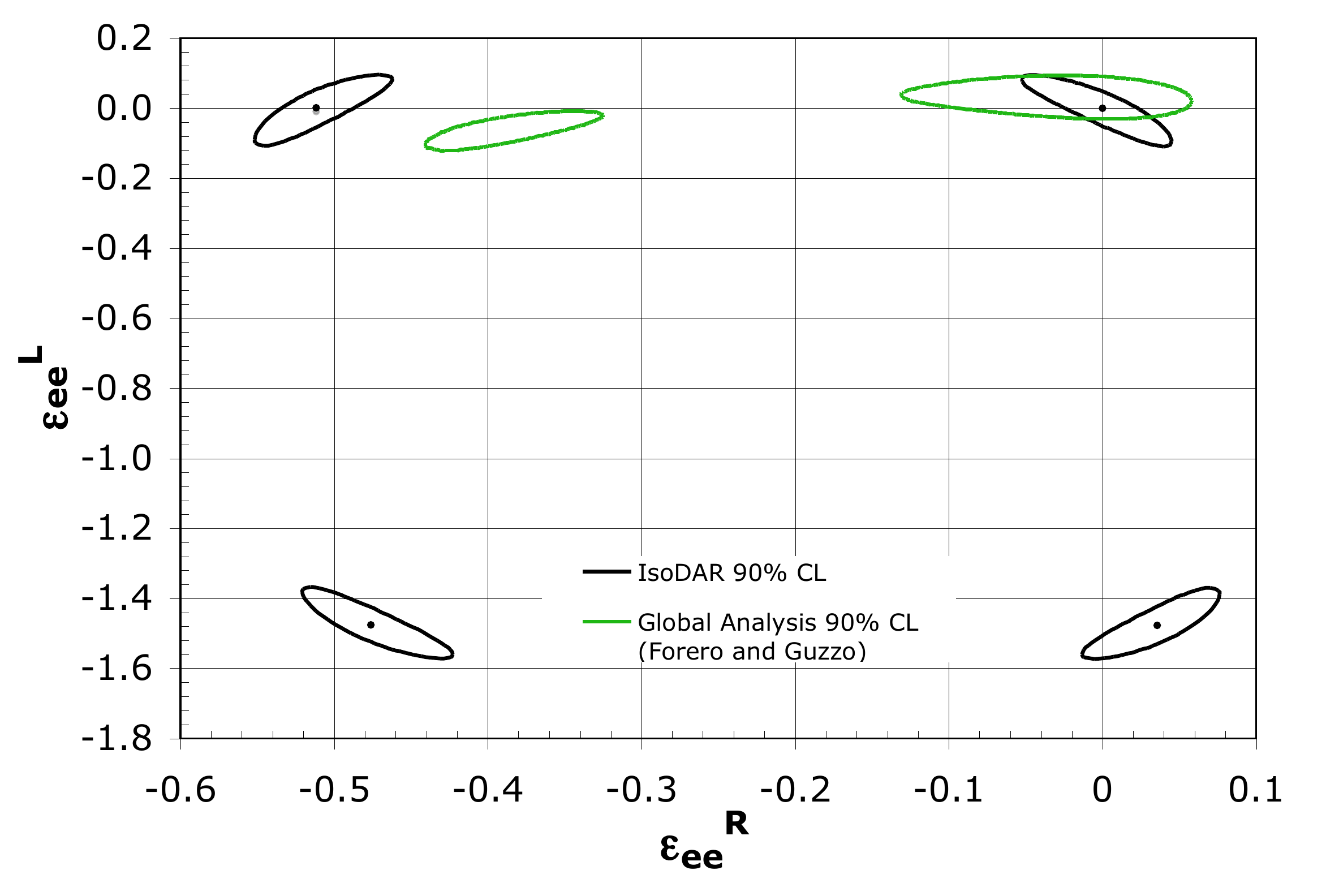}
		\includegraphics[width=0.49\textwidth]{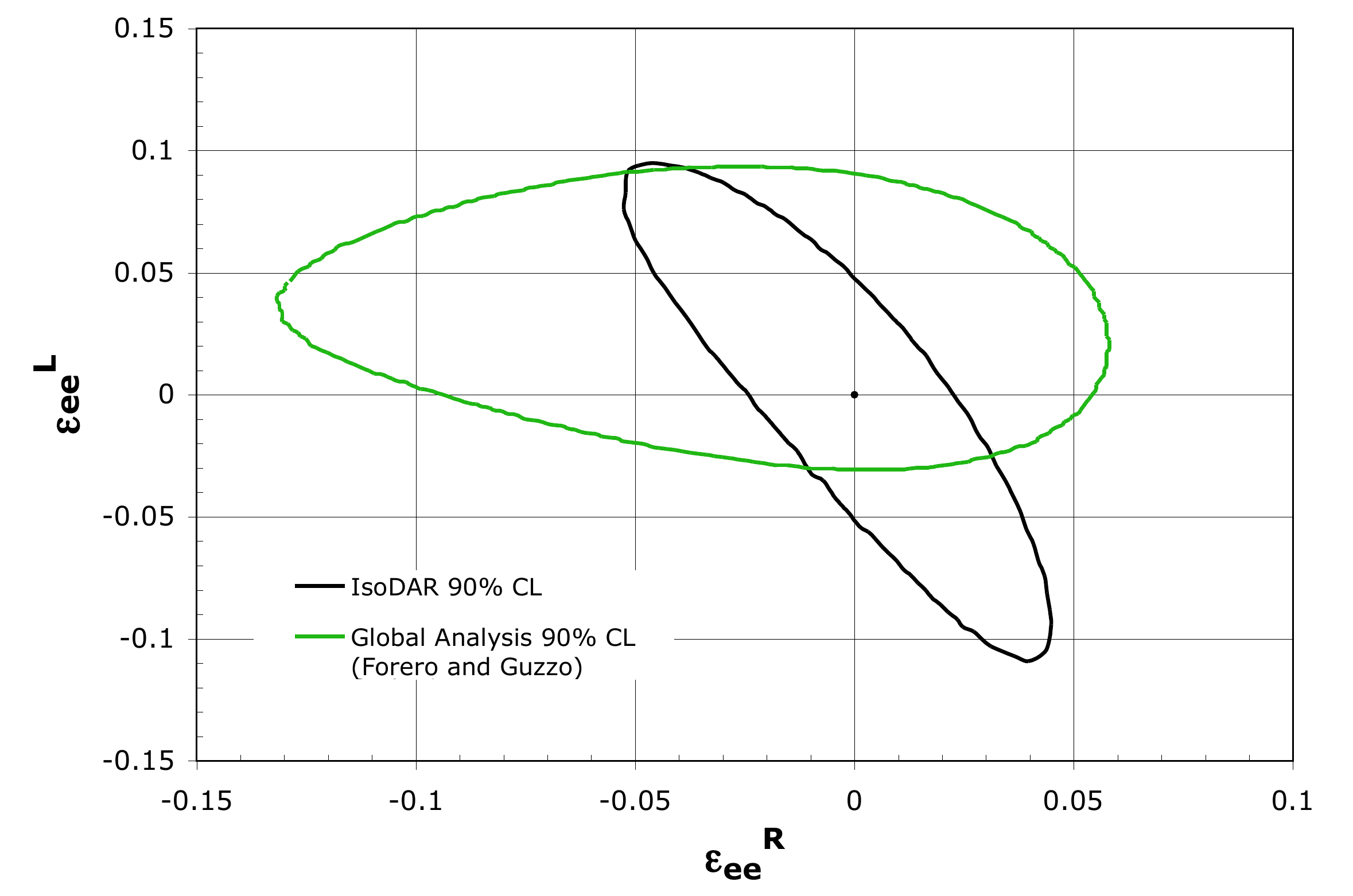}
	\end{center}
	\caption{\label{NSI} (Top) IsoDAR's sensitivity to $\epsilon_{e e}^{e L}$ and $\epsilon_{e e}^{e R}$. The current global allowed region, based on Ref.~\cite{NSI_forero_guzzo} is also shown. (Bottom) A zoomed-in version of the top plot, emphasizing the region near $\epsilon_{ee}^{e  L}$ and $\epsilon_{ee}^{e  R}$ $\sim0$ is shown.}
\end{figure}

Table~\ref{events} shows the expected signal and background event totals assuming a nominal 5 year IsoDAR run with a 90\% duty factor.  We assume that the energy spectrum of the non-beam backgrounds can be measured with 4.5 years of KamLAND data before the IsoDAR source turns on.  The energy spectrum of the non-beam background, mis-identified IBD events, can be extracted from beam-on data with a dedicated delayed coincidence selection.  Given these assumptions, Table~\ref{results} gives the IsoDAR sensitivity to $\sin^2\theta_W$ from a combined fit to the rate and spectral ``shape'' of the differential ES cross section, as well as each individually.  Sensitivities are also shown for the case of a 50\% background reduction and for the case of a 100\% background reduction.

To compare the sensitivity of IsoDAR with that of other experiments, the fits to the ES cross section can also be done in terms of $g_V$ and $g_A$. Fig~\ref{gV_gA} shows the IsoDAR 1$\sigma$ contour in  the $g_V$--$g_A$ plane as well as contours from other experiments. IsoDAR would be the most sensitive $\nu_e e/\bar\nu_e e$ experiment to date and could test the consistency of $\nu_e e/\bar\nu_e e$ couplings with $\nu_\mu e/\bar\nu_\mu e$ couplings.

Finally, we can also estimate IsoDAR's sensitivity to the non-universal NSI parameters $\epsilon_{ee}^{e L}$ and $\epsilon_{ee}^{e R}$, assuming the Standard Model value for $\sin^2\theta_W=0.238$.  The results are shown in Fig.~\ref{NSI} along with the current global allowed region~\cite{NSI_forero_guzzo}.  In the region around $\epsilon_{ee}^{e L}$ and $\epsilon_{ee}^{e R}\sim0$, the IsoDAR 90\% confidence interval significantly improves the global picture.

\subsection{Coherent neutrino scattering at IsoDAR}
As discussed in the previous section, an intense source of neutrinos provides an immense opportunity for a number of physics measurements other than a sterile neutrino search. Along with the weak mixing angle measurement and sensitivity to non-standard neutrino interactions, such a source could allow the first detection of and
subsequent high statistics sampling of coherent, neutrino-nucleus scattering events. Although the process is 
well predicted by the Standard Model and has a comparatively large cross section in the
relevant energy region ($\sim$10-15 MeV), neutral current coherent scattering has never been observed before 
as the low energy nuclear recoil signature is difficult to observe.

A modest sample of a few hundred events collected with a
keV-scale-sensitive dark matter style detector could improve upon
existing non-standard neutrino interaction parameter sensitivities by
an order of magnitude or more. A deviation from the $\sim$5\% predicted
cross section could be an indication of new physics. Furthermore, the
cross section is relevant for understanding the evolution of core
collapse supernovae as well as for characterizing future burst supernova
neutrino events collected with terrestrial detectors.

A dark matter style detector with keV-scale sensitivity to nuclear recoil events, perhaps based on 
germanium crystal or single phase liquid argon/neon technology, in combination with an intense proton
source such as IsoDAR could perform the physics discussed above. The technology currently exists for such 
a detector (with requisite passive and active shielding) to be deployed on the surface or underground. Note that
the KamLAND detector is not sensitive enough for such a measurement.

\begin{figure}[tb]
\begin{center}
\begin{tabular}{c c}
\includegraphics[scale=.42]{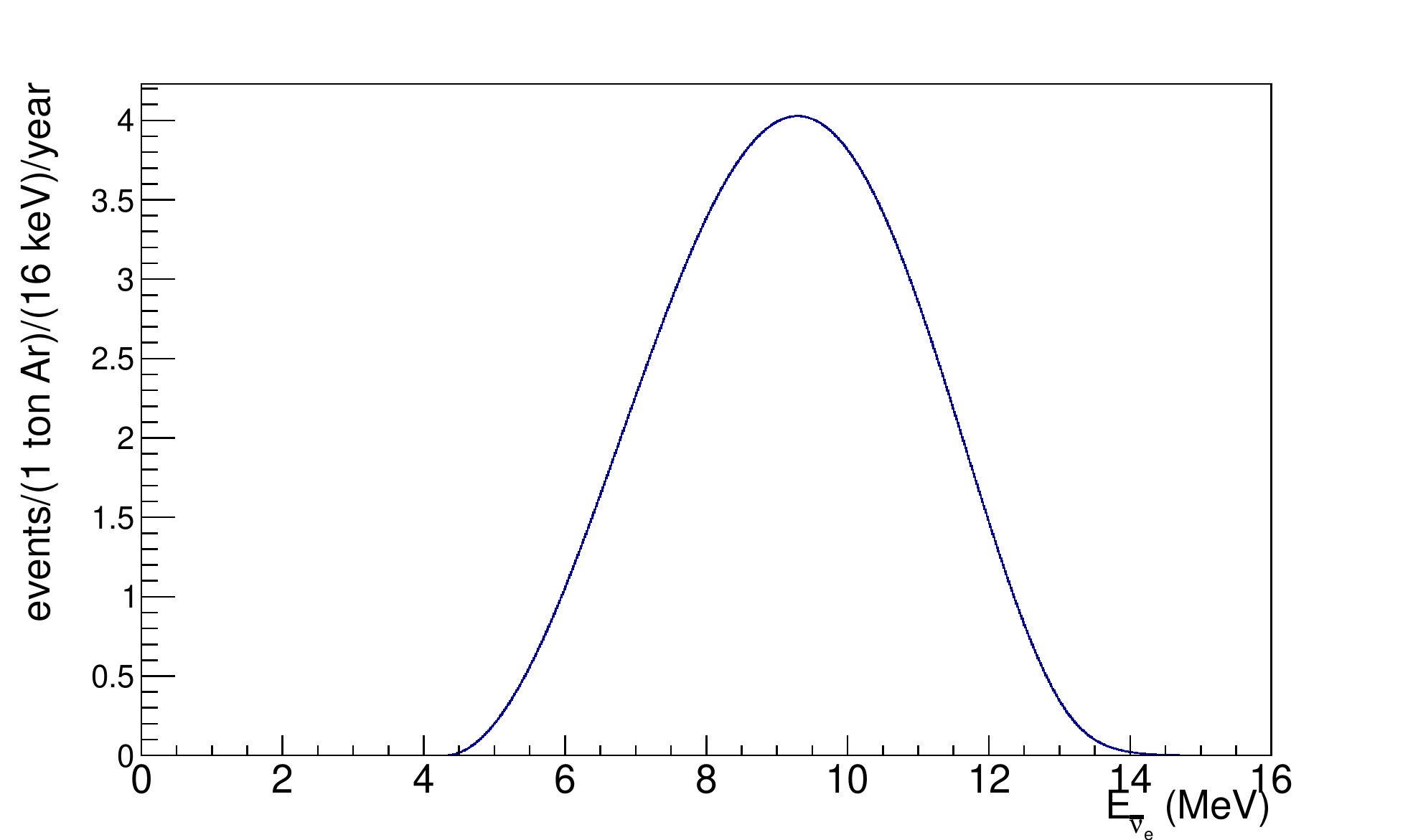}
&
\includegraphics[scale=.42]{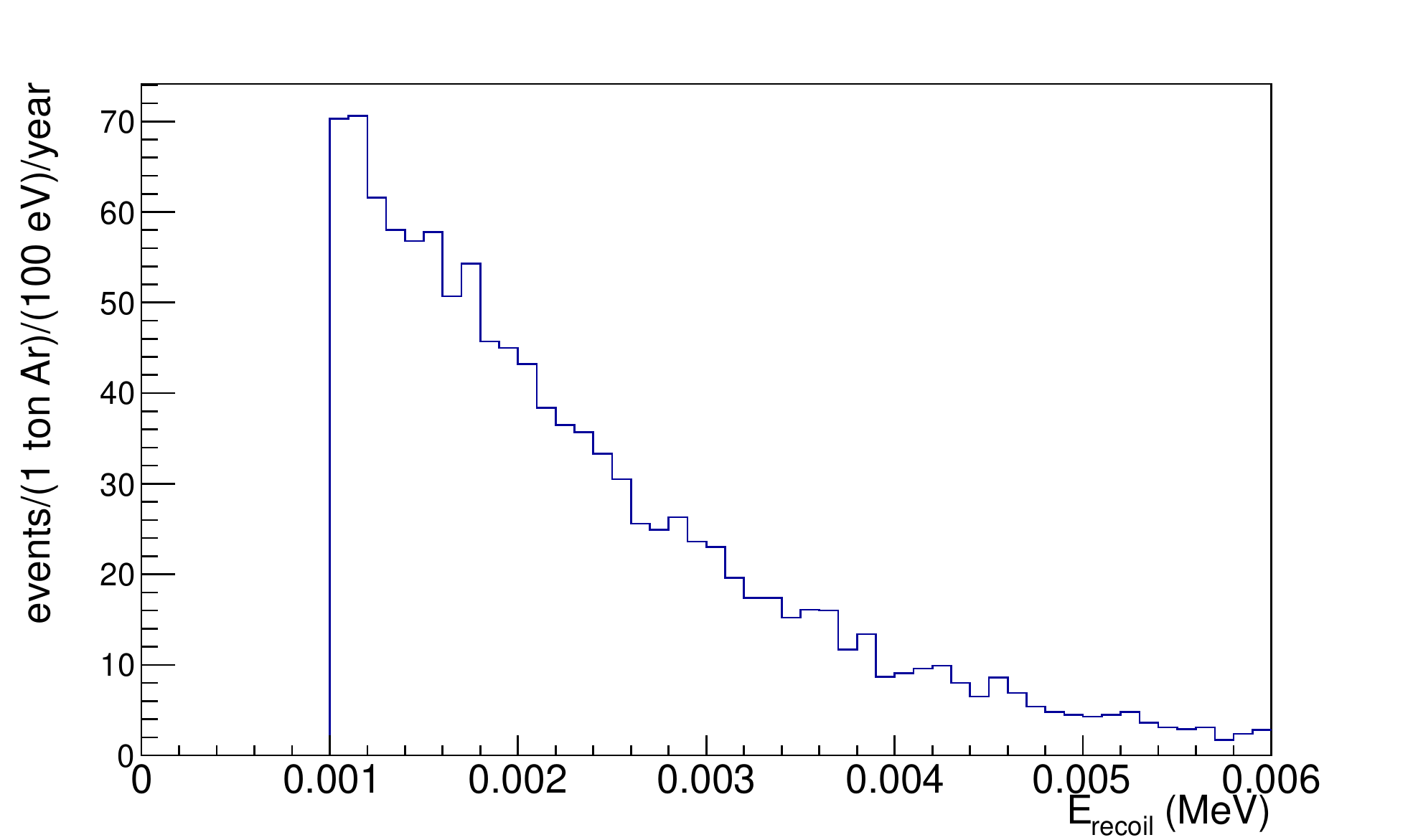}
\end{tabular}
\end{center}
\vspace{-.5cm}
\caption{(Left) Coherent event rate in terms of antineutrino energy with a 1000~kg argon detector at a 10~m average baseline from the IsoDAR source. (Right) The event rate in terms of nuclear recoil energy.}
\label{neutrinoE_coherent}
\end{figure}

Figure~\ref{neutrinoE_coherent} shows the expected rates in terms of neutrino energy and nuclear recoil energy for an IsoDAR source
(2.58$\times10^{22}~\overline{\nu}_e$/year) in combination with a 1000~kg argon detector at a
10~m average baseline from the source with a 1~keV nuclear recoil energy threshold and 20\% energy resolution. 
Given these assumptions, about 1200~events per year could be collected for a high statistics
sampling of this event class. A first observation of the process is clearly possible with a more modest size detector as well.

\section{DAE$\delta$ALUS}
\subsection{$CP$ violation searches}
$CP$ violation can occur in neutrino oscillations  if there is a complex phase, $\delta_{CP}$, in
the $3\times3$ neutrino mixing matrix between the neutrino flavor and mass
eigenstates.  Observation of CP violation in the light neutrino sector would  be a first hint of such effects 
in the early universe where GUT-scale Majorana neutrinos can have $CP$ violating decays that lead to
the matter-antimatter asymmetry that we now observe.  This process is called ``leptogenesis.''~\cite{Murayama:1993em, Ma:1998dx, Davidson:2002qv}

The parameter $\delta_{CP}$ is accessible through the muon-to-electron
neutrino flavor oscillation probability.  For oscillations in a vacuum,  the probability is given by
\cite{Nunokawa:2007qh}:
\begin{align}
P_{\mu\rightarrow e} & =\sin^{2}\theta_{23}\sin^{2}2\theta_{13}\sin^{2}%
\Delta_{31} \nonumber\\
& \mp\sin\delta\sin2\theta_{13}\sin2\theta_{23}\sin2\theta_{12}\sin^{2}%
\Delta_{31}\sin\Delta_{21} \nonumber \\
& +\cos\delta\sin2\theta_{13}\sin2\theta_{23}\sin2\theta_{12}\sin\Delta
_{31}\cos\Delta_{31}\sin\Delta_{21} \nonumber\\
& +\cos^{2}\theta_{23}\sin^{2}2\theta_{12}\sin^{2}\Delta_{21} 
\label{equ:beam}
\end{align}
where $\Delta_{ij}=\Delta m_{ij}^{2}L/4E_{\nu}$.  In the second term,
the $-(+)$ refers to neutrino (antineutrino) operation.  

A critical parameter for measuring $CP$ violation is the size of the mixing
angle $\theta_{13}$, which determines the size of the first three terms in
Eq.~\ref{equ:beam}.  Recently, several reactor neutrino disappearance 
experiments (Double Chooz, Daya Bay, and RENO) have made precision
measurements of $\theta_{13}$ giving a global average of 
$\theta_{13} = 8.75^\circ \pm 0.43^\circ$~\cite{GonzalezGarcia:2012sz}.  
The fact that $\theta_{13}$ is now known to be fairly large makes the
search for $CP$ violation viable and a key next step in particle physics.

For long baseline experiments, searches for $CP$ violation rely on
comparing neutrino and antineutrino oscillation probabilities, thus,
exploiting the above change of sign in order to isolate $\delta_{CP}$.  This type
of measurement is complicated by matter effects, in which the 
forward scattering amplitude for neutrinos and antineutrinos
differs due to the presence of electrons, rather than positrons, in matter.  The
matter effects result in a modification of Eq.~\ref{equ:beam} giving

\begin{eqnarray}
P_{\mu\rightarrow e}^{matter}  = 
&&\sin^{2}\theta_{23}\sin^{2}2\theta_{13}\frac{\sin
^{2}\left(  \Delta_{31}\mp aL\right)  }{\left(  \Delta_{31}\mp aL\right)
^{2}}\Delta_{31}^{2} \nonumber \\
&& \mp\sin\delta\sin2\theta_{13}\sin2\theta_{23}\sin2\theta_{12}\sin\Delta
_{31}\frac{\sin\left(  \Delta_{31}\mp aL\right)  }{\left(  \Delta_{31}\mp
aL\right)  }\Delta_{31}\frac{\sin\left(  aL\right)  }{\left(  aL\right)
}\Delta_{21} \nonumber \\
&& +\cos\delta\sin2\theta_{13}\sin2\theta_{23}\sin2\theta_{12}\cos\Delta
_{31}\frac{\sin\left(  \Delta_{31}\mp aL\right)  }{\left(  \Delta_{31}\mp
aL\right)  }\Delta_{31}\frac{\sin\left(  aL\right)  }{\left(  aL\right)
}\Delta_{21} \nonumber\\
&& +\cos^{2}\theta_{23}\sin^{2}2\theta_{12}\frac{\sin^{2}\left(  aL\right)
}{\left(  aL\right)  ^{2}}\Delta_{21}^{2}  \label{Pmatter}.
  \end{eqnarray}
In this equation, $a=\frac{G_{F}N_{e}}{\sqrt{2}}$ 
and  $\mp$ refers to neutrinos
  (antineutrinos).  
  Matter effects only appear when $L$ is large, because $a \approx (
  3500\text{ km}) ^{-1}$ (with $\rho
  Y_{e}=3.0\text{ g/cm}^{3}$) is small.  Short-baseline experiments, such
  as DAE$\delta$ALUS and moderate baseline experiments such as T2K
  at $L = 295$ km suffer negligible matter effects.  On the other hand,
  long baseline experiments such as NO$\nu$A and LBNE have
  significant matter effects \cite{LBNEPX}.
 The terms that are modified by the matter
  effects also depend on $\mathop{\mathrm{sign}}(\Delta m_{31}^2)$,
  making the corrections dependent on knowing this sign, commonly
  called the ``mass hierarchy.''
  
\begin{figure}[tb]
\begin{center}
\includegraphics[scale=.47]{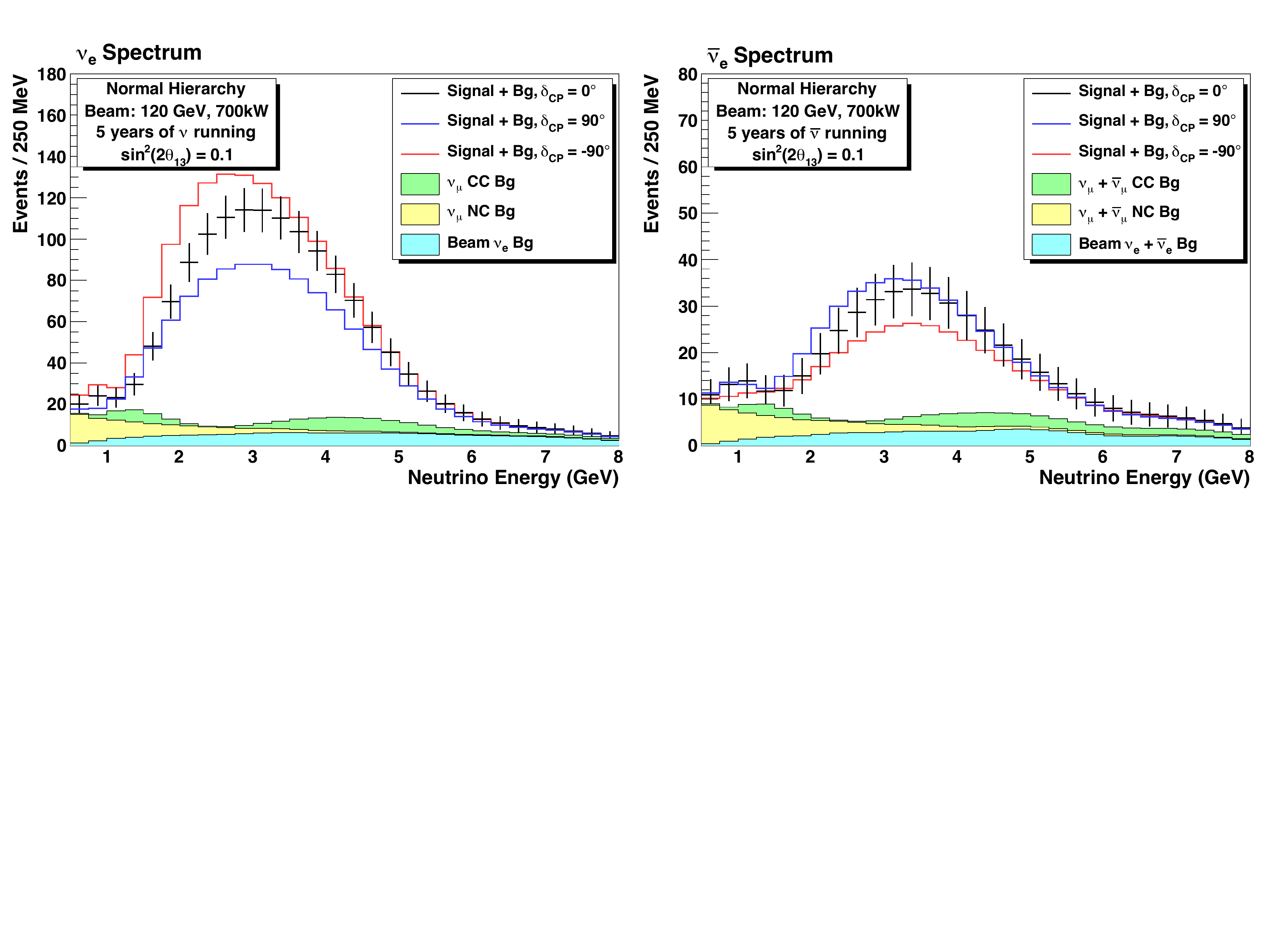}
\end{center}
\vspace{-6.5cm}
\caption{Estimated events for the LBNE experiment running for 5 years neutrino
and 5 years antineutrino with 700 kW of 120 GeV protons on target and with the  assumption that $\sin^2 2\theta = 0.1$.  Backgrounds
and changes with $\delta_{CP}$ are also shown. These plots are from Ref.~\cite{PWG_report}.
\label{LBNE_events}}
\end{figure}

  For long baseline accelerator oscillation experiments, gathering sufficient 
  antineutrino data sets is difficult due to the reduced negative pion production
  rate by accelerator protons and by the reduced interaction cross section of
  antineutrinos.  The current event estimates for the LBNE experiment with a 34~kton
  liquid argon detector at 1300~km from the Fermilab site is shown in 
  Figure~\ref{LBNE_events} \cite{LBNEPX} for the normal mass hierarchy.

In contrast, the DAE$\delta$ALUS experiment will be a search only in the antineutrino mode $\bar \nu_\mu \rightarrow \bar \nu_e$ with no matter effects, and with reduced backgrounds and systematic uncertainties, plus a unique experimental layout in which several low-cost neutrino sources are at different distances from one large detector.  With an antineutrino-only beam, the oscillation probability is given by Eq.~\ref{equ:beam}, and the sensitivity to $CP$ violation comes about through the
interference between $\Delta_{12}$ and $\Delta_{13}$ transitions, which have a distinctive $L$ dependence.  

Specifically, DAE$\delta$ALUS will search for $\bar \nu_\mu \rightarrow \bar \nu_e$
oscillations using neutrinos from three stopped-pion DAR sources, which interact in a single large 200 to 500~kton Gd-doped water Cherenkov or a large 50~kton liquid scintillator detector.  The spectrum of $\bar\nu_\mu$'s that can oscillate into $\bar\nu_e$'s is shown in Figure~\ref{flux_xsec}.  The detection of the electron antineutrinos is done through the IBD process where the outgoing positron is once again required to have a delayed coincidence with a neutron capture on Gd for the water detector or on hydrogen for the scintillator detector.  This process has a high cross section at $\approx 50$~MeV, but requires either Gd doping for a water detector or a scintillator detector to detect the outgoing neutron and separate the IBD events from the preponderance of charged-current $\nu_e$ events.
 
\begin{figure}[tb]
\begin{center}
{\includegraphics[width=4.5in]{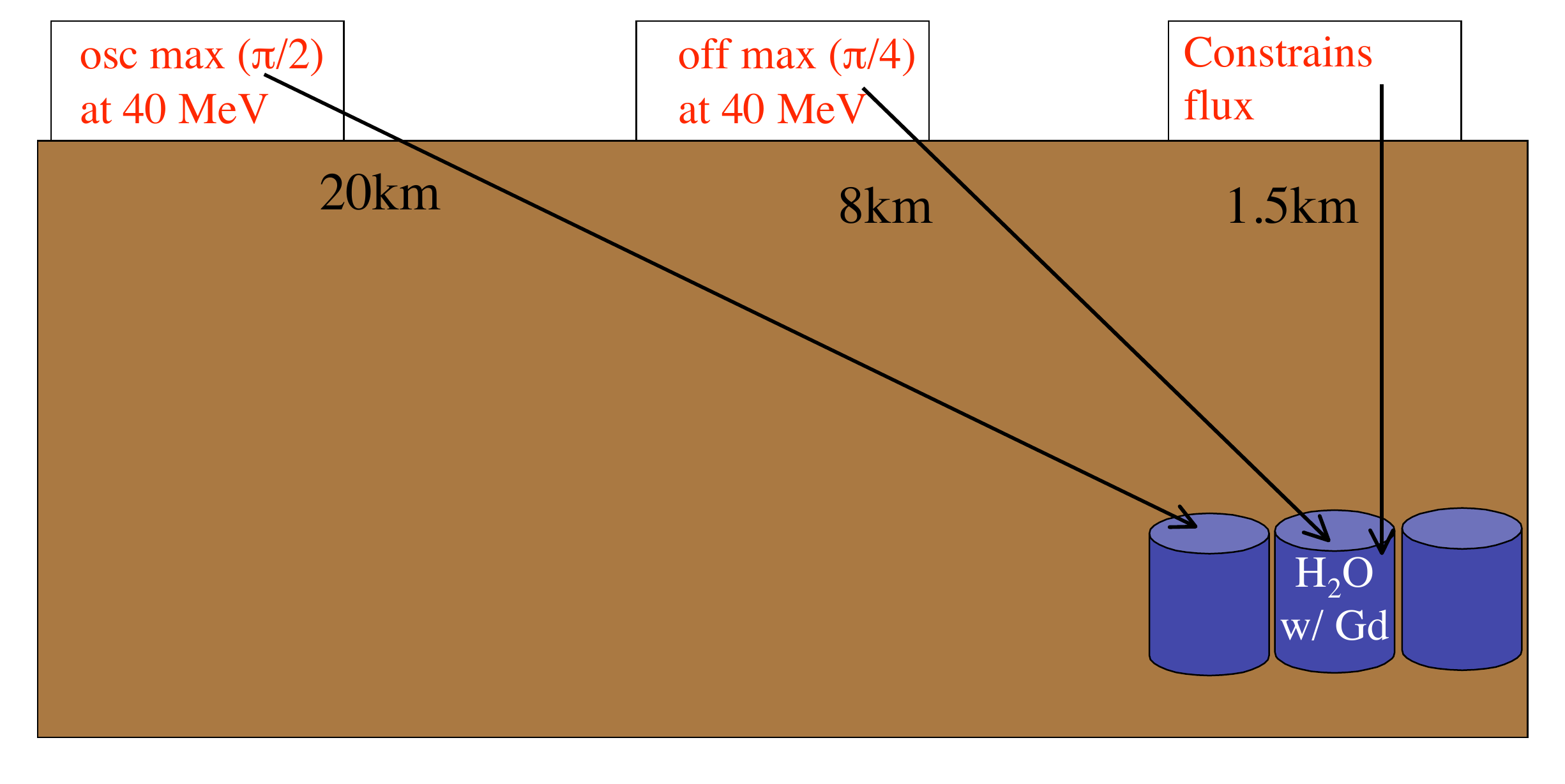}
} \end{center}
\vspace{-0.25in}
\caption{Schematic of the DAE$\delta$ALUS experiment.  Three 
neutrino source locations are used in conjunction with a large water 
Cherenkov or scintillator based detector. 
\label{schematic} }
\end{figure}

The accelerators will be positioned at 1.5, 8, and 20~km from the large detector as shown in Figure~\ref{schematic} and are all above ground to reduce the installation and running complexity.  Each accelerator provides different physics data for the $CP$ violation search.  The 1.5-km accelerator allows measurement of the
beam-on backgrounds and the normalization.  The 8~km site is at an oscillation wavelength of about $\pi/4$ at 50~MeV and the 20~km site is at oscillation maximum for this energy.  Each site will be run for 20\% of the time so that the events from a given source distance can be identified by their time-stamp with respect to this running cycle.   This will leave 40\% of the time for beam-off running to measure the non-beam backgrounds and provide other physics data.   The baseline plan is for a ten~year run with 1~MW, 2~MW, and 5~MW neutrino sources at the 1.5, 8, and 20~km sites, respectively.

Combining the data from the three accelerators helps to minimize the  systematic uncertainties associated with the beam and detector and leads to a highly sensitive search for $CP$ violation.  The shape of the DAR flux with energy is known to high precision and is common among the various distances; thus shape comparisons will have small uncertainties.  The interaction and detector systematic errors are low since all events are detected in a single detector.  The fiducial volume error on the IBD events is also small due to the extreme volume-to-surface-area ratio of the ultra-large detector.  Therefore, the main errors for the measurements are related to the statistics of the data and to normalization uncertainties.  The normalization uncertainties are dominated by the neutron tagging efficiency, assumed to be 0.5\%, and the antineutrino flux uncertainties that are constrained as described next.

The DAE$\delta$ALUS $CP$ violation analysis follows three steps.  First, the absolute normalization of the flux from the near accelerator is measured using the $>$21,000 neutrino-electron scatters from that source in the detector, for which the cross section is known to 1\%.  The relative flux normalization between the sources is then determined using the comparative rates of charged current $\nu_e$-oxygen (or $\nu_e$-carbon) interactions in the the detector.  Since this is a relative measurement, the cross section uncertainty does not come in but the high statistics is important.  Once the normalizations of the accelerators are known, then the IBD data can be fit to extract the $CP$-violating parameter $\delta_{CP}$.  The fit needs to include all the above systematic uncertainties along with the physics parameter uncertainties associated with, for example, the knowledge of $\sin^22\theta_{13}$ and $\sin^2\theta_{23}$, which are assumed to be known with an error of $\pm 0.005$ and $\pm 0.01$, respectively.

DAE$\delta$ALUS must be paired with water or scintillator detectors
that have free proton targets. The original case was developed for
a 300 kt Gd doped water detector at Homestake, in
coordination with LBNE \cite{config}.   Subsequently,  DAE$\delta$ALUS was
incorporated into a programa with the 50 kt LENA detector \cite{LENA}
(called ``DAE$\delta$ALUS@LENA'').  
This paper introduces a new study, where DAE$\delta$ALUS is paired
with the Gd-doped 560 kt Hyper-K \cite{hyperk} (``DAE$\delta$ALUS@Hyper-K'').  
This results in inprecedented sensitivity to $CP$ violation when
``DAE$\delta$ALUS@Hyper-K'' data is combined with data from  Hyper-K running with the 
750 kW JPARC beam. (``DAE$\delta$ALUS/JPARC@Hyper-K'').  
In this scenario, 
JPARC provides a pure $\nu_\mu$ flux,  rather than running in neutrino
and antineutrino mode.   This plays to the strength of the
JPARC conventional beam, while
DAE$\delta$ALUS provides a high statistics $\bar \nu_\mu$ flux with no $\nu_\mu$ contamination.
A summary of the assumptions for the various
configuration scenarios is provided in Table~\ref{tab:configs}.

\begin{table}[t]
\centering {\footnotesize
\begin{tabular}{|l|c|c|c|c|c|}
\hline
Configuration  &  Source(s) & Average& Detector & 
Fiducial & Run \\ 
Name  &   & Long Baseline &  &  
Volume & Length \\  
 &   & Beam Power &  &  
 &  \\ \hline  
DAE$\delta$ALUS@LENA & DAE$\delta$ALUS only &  N/A & LENA & 50 kt & 10
years \\   \hline
DAE$\delta$ALUS@Hyper-K & DAE$\delta$ALUS only &  N/A & Hyper-K & 560 kt & 10
years\\  \hline
DAE$\delta$ALUS/JPARC & DAE$\delta$ALUS  &   & Hyper-K & 560 kt & 10 
years\\ 
(nu only)@Hyper-K& \& JPARC  & 750 kW  &  & & \\  \hline
JPARC@Hyper-K & JPARC & 750 kW & Hyper-K & 560 kt & 3 years $\nu$ +\\
 & &  && & 7 
years $\bar \nu$ \cite{hyperk}\\  \hline
LBNE & FNAL & 850 kW & LBNE & 35 kt & 5 years $\nu$ \\
         &          &              &          &          &  5 years
         $\bar \nu$ \cite{LBNEPX}\\  
\hline
\end{tabular}}
\caption{\it \footnotesize  Configurations considered in the various
$CP$ violation sensitivity studies\label{tab:configs}. }
\end{table} 

\begin{figure}[htb]
\begin{center}
\begin{tabular}{c c}
\includegraphics[width=3.3in]{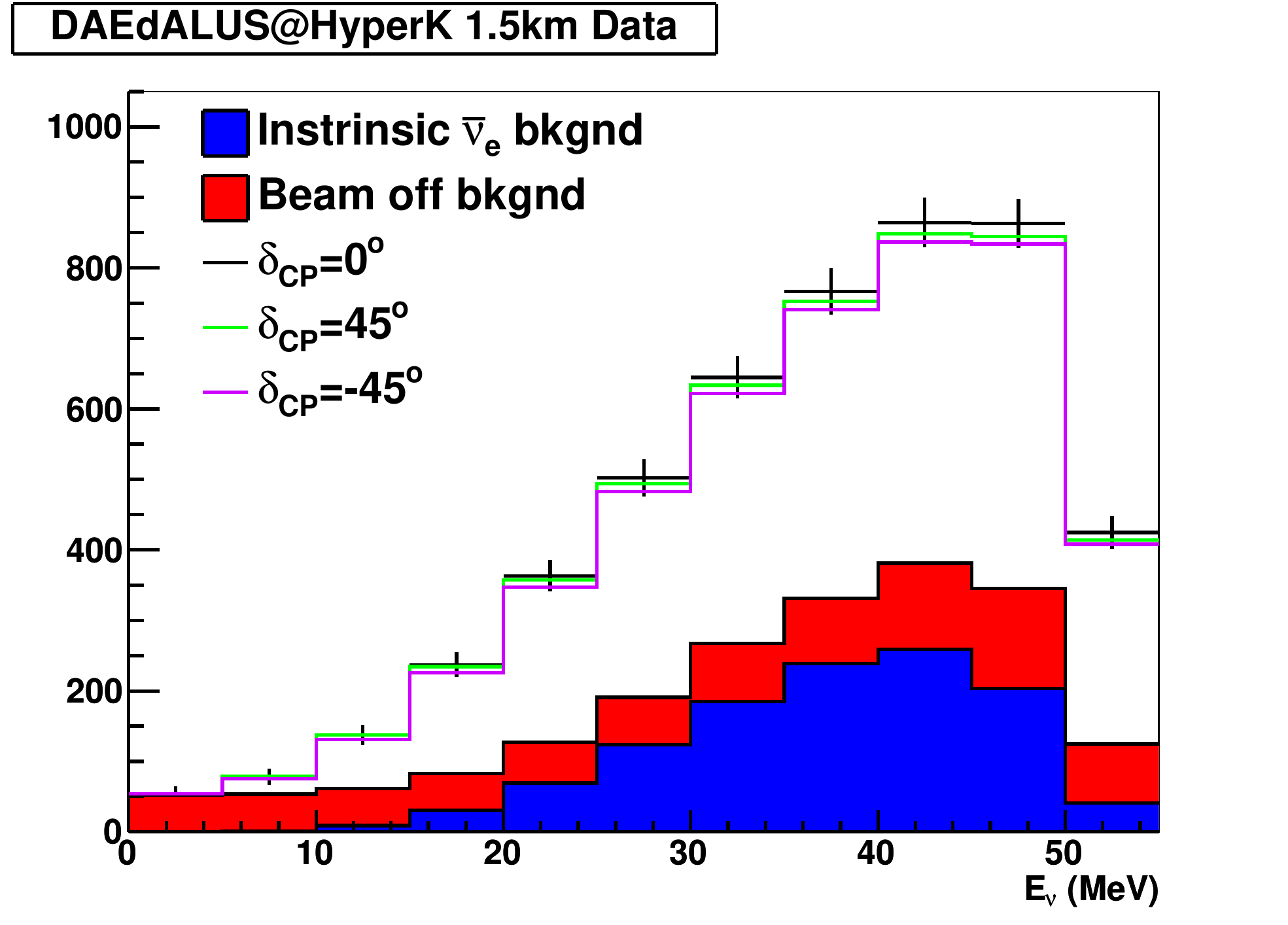}
&
\includegraphics[width=3.3in]{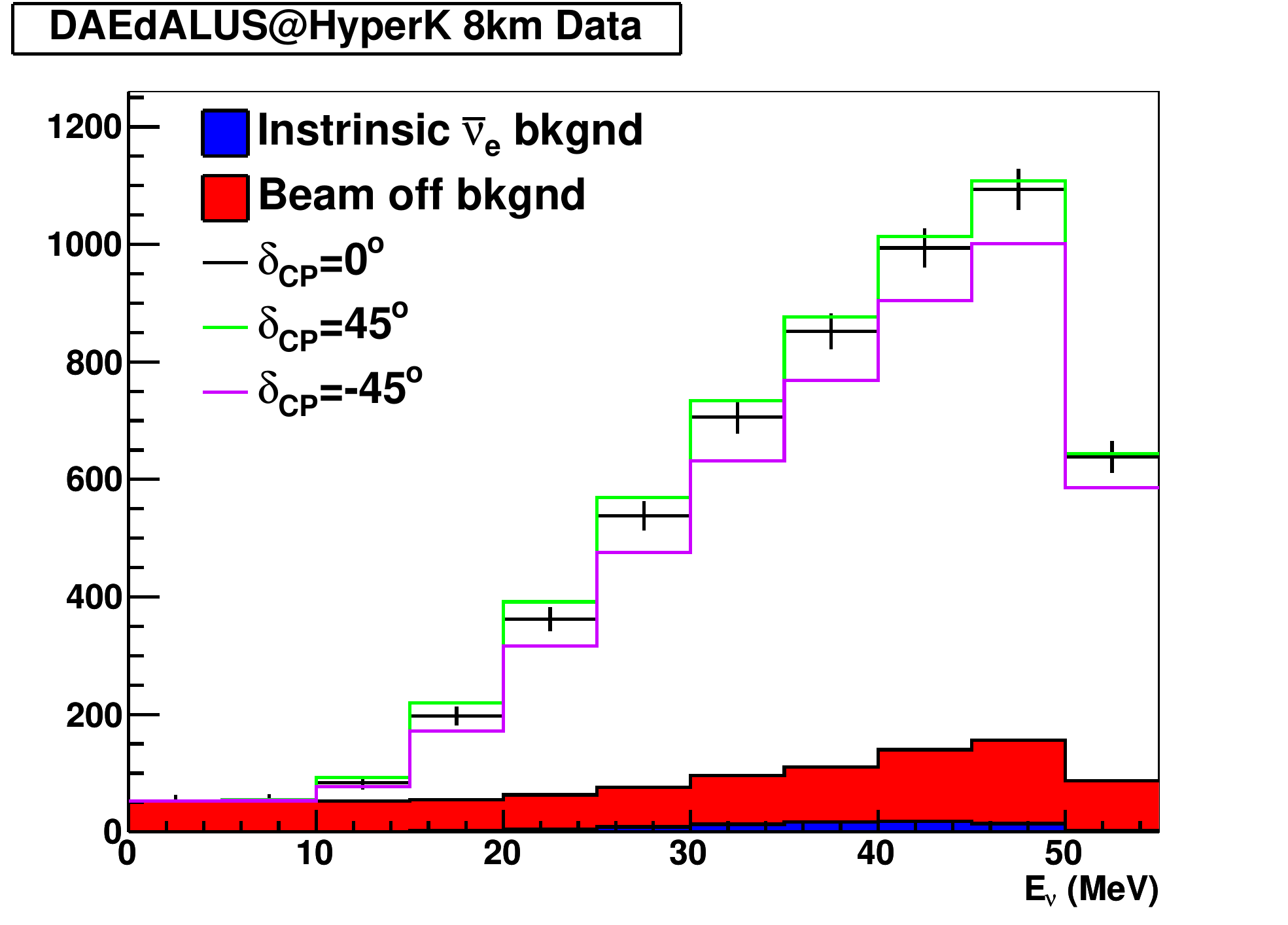}
\end{tabular}
\\
\includegraphics[width=3.3in]{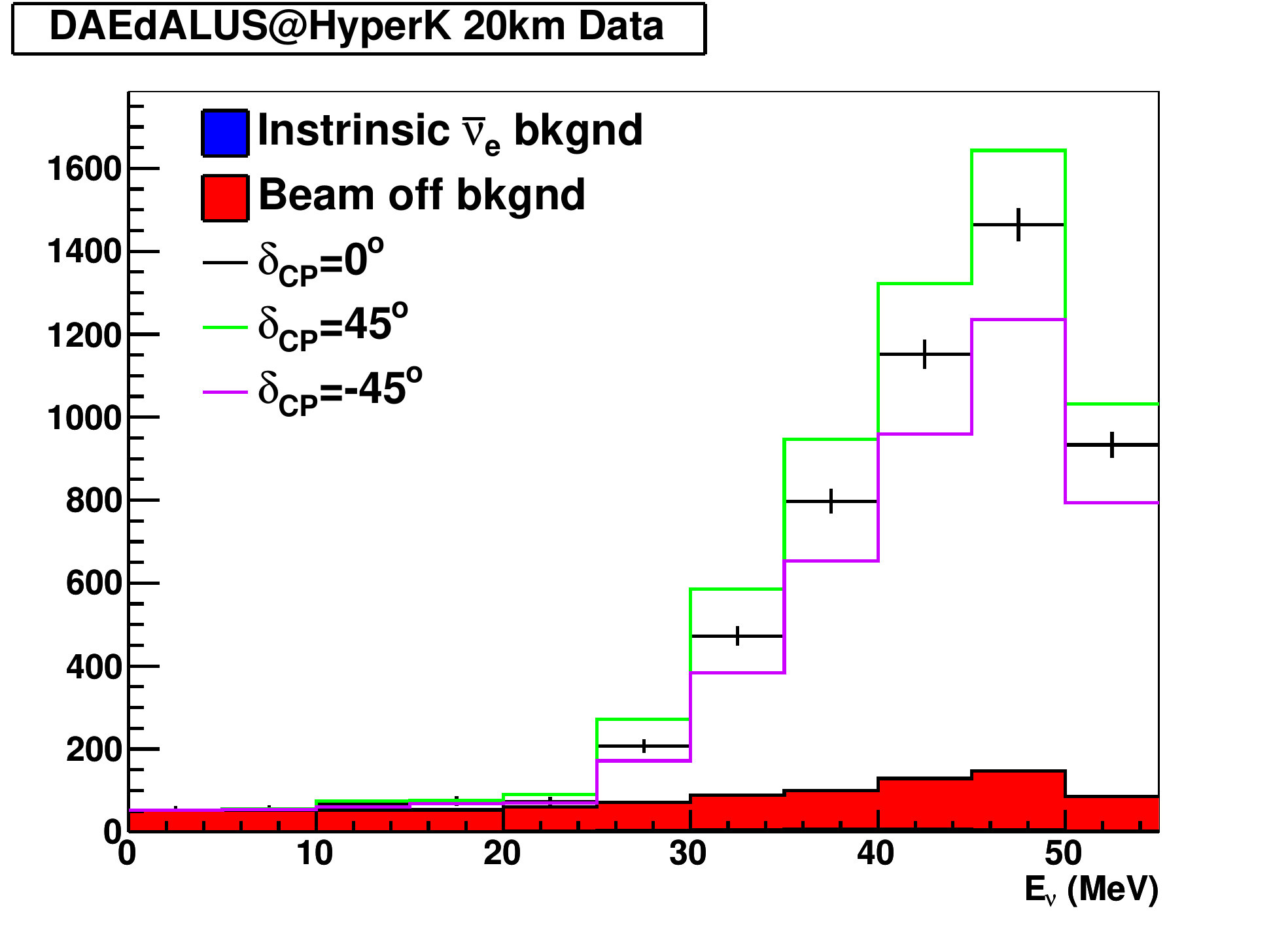}
\end{center}
\vspace{-.6cm}
\caption{The event energy distributions for signal and background 
of the DAE$\delta$ALUS@Hyper-K  running scenario
with $\sin^2 2\theta_{13}=0.10.$.   Black, green and violet histograms
show signals for $\delta_{cp}=0$, 45$^\circ$ and -45$^\circ$.  The
blue histogram shows the intrinsic $\bar \nu_e$ beam-on background. The
red histogram shows the beam-off backgrounds.   Top row:  events from the near (1.5 km) 
and middle (8 km) accelerators.   Bottom row: events from the far (20 km) accelerator.
\label{eventsbkgds}}
\end{figure}

\begin{figure}[htbp]
\vspace{-1.0in}
\begin{center}
{\includegraphics[width=5.in]{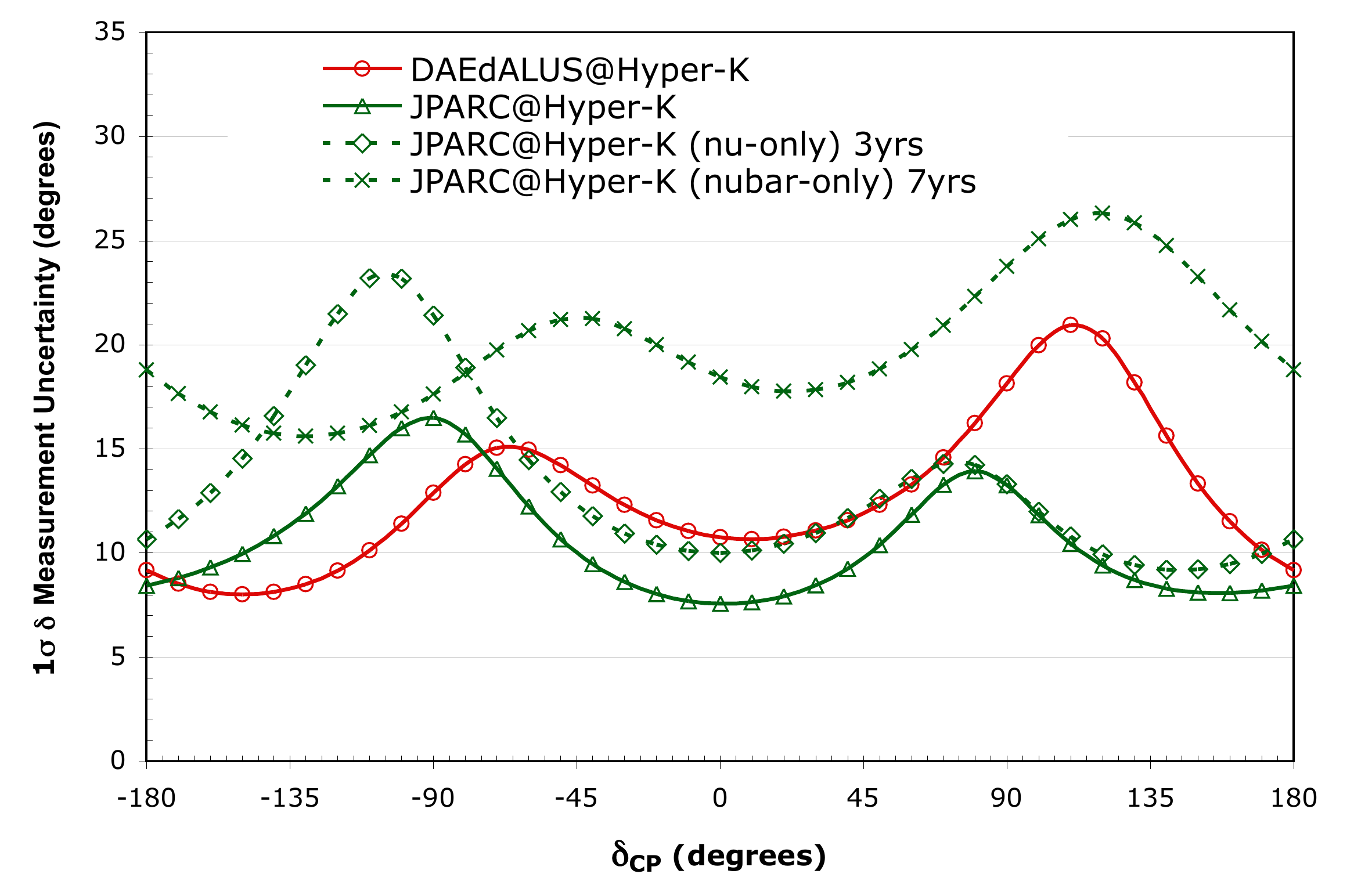}}\\
{\includegraphics[width=5.in]{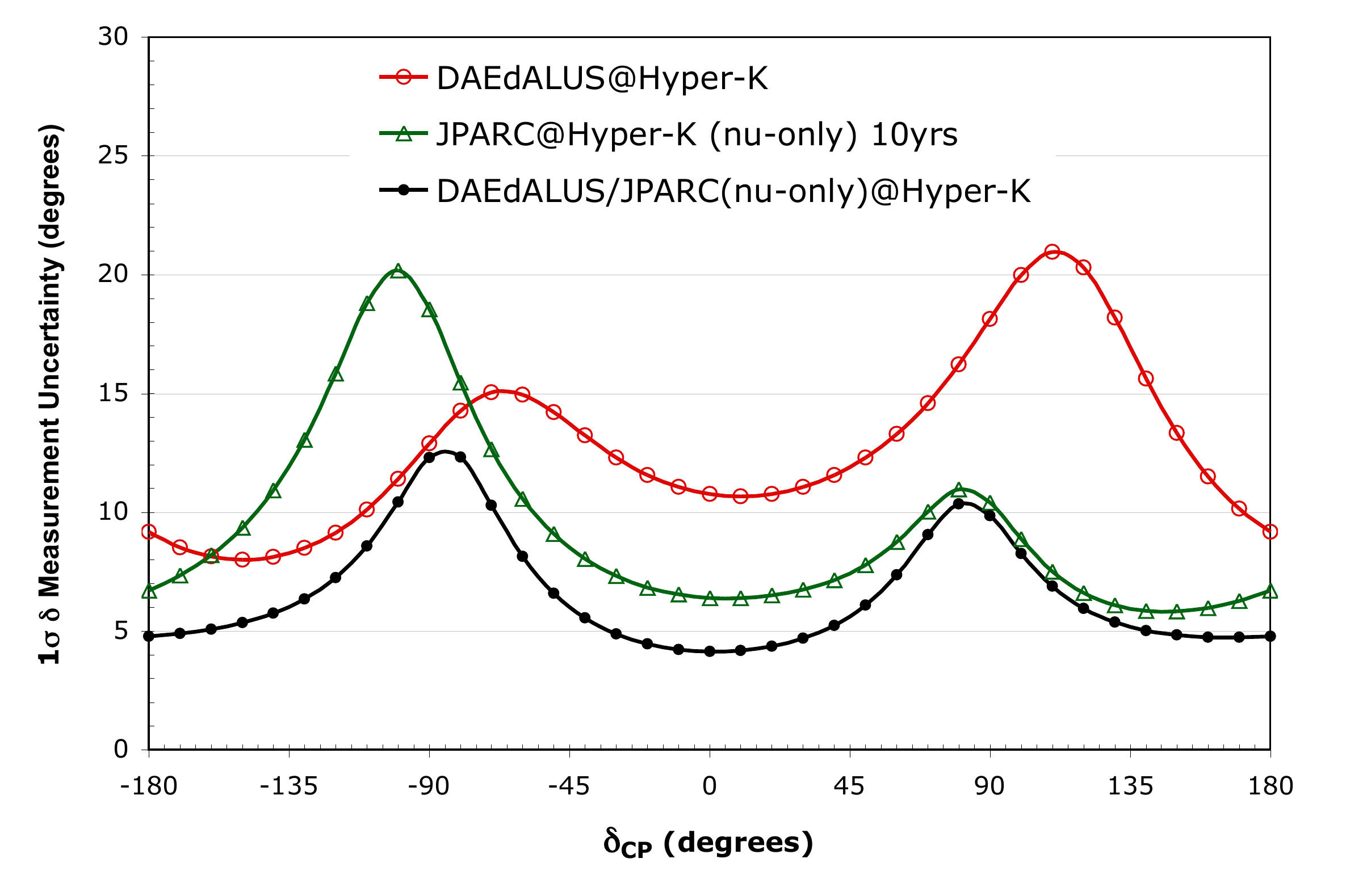}}
\end{center}
\vspace{-0.3in}
\caption{ 
\label{cp_error}   Top:   1 $\sigma$ measurement sensitivities 
for the nominal JPARC@Hyper-K run (solid
green with triangles) compared to DAE$\delta$ALUS@Hyper-K (sold red).
Dashed green curves show contributions of neutrino and antineutrino 
running to the total sensitivity of JPARC@Hyper-K .   Bottom:  
DAE$\delta$ALUS/JPARC(nu only)@Hyper-K combined measurement sensitivity.   
The contribution to the combined measurement from
DAE$\delta$ALUS antineutrinos is indicated in red and the contribution from  
JPARC neutrinos is indicated in green.}
\end{figure}

$CP$ violation sensitivities have been estimated for 10~year baseline data sets 
for all the configurations given in Table~\ref{tab:configs} using a 
$\Delta \chi^2$ fit with pull parameters for each of the systematic uncertainties.  
For the DAE$\delta$ALUS configurations, data from all three neutrino sources are included along with the neutrino-electron and $\nu_e$-oxygen (or $\nu_e$-carbon) normalization samples.  As an example, Table~\ref{events_123_127_HK} 
and Figure~\ref{eventsbkgds} presents a summary of the events by category for the
DAE$\delta$ALUS@Hyper-K configuration.
The precision for measuring the $\delta_{CP}$ parameter in the 
DAE$\delta$ALUS@Hyper-K configuration is given in Table~\ref{deltaCP_123_127} for 
$\sin^22\theta_{13} = 0.10$ \cite{GonzalezGarcia:2012sz}, both for the total and statistical-only uncertainty. The distribution of the uncertainty as a function of $\delta_{CP}$ is shown in Figure~\ref{cp_error}.  From these estimates, it is clear that, even with the large Hyper-K detector, the measurement is dominated by statistical uncertainty.   Also shown in the table are estimates of the measurement uncertainties for the proposed Hyper-K~\cite{hyperk} and LBNE~\cite{LBNE} experiments for ten~year runs with the proposed upgraded beam intensities (0.75~MW for HyperK and an average 0.85~MW for LBNE).  Depending on the true value, DAE$\delta$ALUS has comparable sensitivity for measuring $\delta_{CP}$ but has very different systematic uncertainties.  Thus, DAE$\delta$ALUS could provide key information that can be used in conjunction with the other experiments to reduce the global measurement uncertainty.

\begin{table}[htb] 
\centering
\begin{tabular}
[c]{l|ccc}\hline
Event Type & 1.5 km & 8 km & 20 km\\\hline
IBD Oscillation Events (E$_{vis}>20$ MeV) &  &  & \\
$\delta_{CP}=0^{0}$, Normal Hierarchy & 2660 &       4456 &       4417\\
\quad\quad" \quad, Inverted Hierarchy & 1838 &       3268 &       4338 \\
$\delta_{CP}=90^{0}$, Normal Hierarchy & 2301 &       4322 &       5506\\
\quad\quad" \quad, Inverted Hierarchy & 2301 &       4328 &       5556\\
$\delta_{CP}=180^{0}$, Normal Hierarchy & 1838 &       3263 &       4295 \\
\quad\quad" \quad, Inverted Hierarchy & 2660 &       4462 &       4460\\
$\delta_{CP}=270^{0}$, Normal Hierarchy & 2197 &       3397 &       3206\\
\quad\quad" \quad, Inverted Hierarchy & 2197 &       3402 &       3242 \\\hline
IBD from Intrinsic $\overline{\nu}_{e}$ (E$_{vis}>20$ MeV) & 1119&79&31\\
IBD Non-Beam (E$_{vis}>20$ MeV) &  &  & \\
\multicolumn{1}{r|}{atmospheric $\nu_{\mu}p$ \textquotedblleft invisible
muons\textquotedblright} & 505 & 505 & 505\\
\multicolumn{1}{r|}{atmospheric IBD} & 103 & 103 & 103\\
\multicolumn{1}{r|}{diffuse SN neutrinos} & 43 & 43 & 43\\\hline
$\nu-$e Elastic (E$_{vis}>10$ MeV) & 40025&2813&1123\\\hline
$\nu_{e}-$oxygen (E$_{vis}>20$ MeV) &188939&13281&5305\\\hline
\end{tabular}
\caption{Event samples for the DAE$\delta$ALUS@Hyper-K  running scenario
for a 10 year run with $\sin^2 2\theta_{13}=0.1$ \cite{GonzalezGarcia:2012sz}.
}\label{events_123_127_HK}%
\end{table}%

\begin{table}[htb] \centering
\begin{tabular}
[c]{l|ccccc}\hline
$\delta_{CP}$ & --180$^\circ$ & --90$^\circ$ & 0$^\circ$ & 90$^\circ$ & 135$^\circ$\\\hline
DAE$\delta$ALUS@Hyper-K  & 9.2 &       12.9 &       10.8 &       18.1 &       16.9\\
\quad Stat - only & (8.8) &     (11.5) &     (10.5) &     (15.8) &     (16.2) \\\hline
JPARC@Hyper-K & 7.8 & 15.2 & 7.8 & 15.0 & 9.1 \\ \hline
LBNE & 10.4 & 18.5 & 10.4 & 15.9 & 11.4 \\ \hline
\end{tabular}
\caption{
DAE$\delta$ALUS@Hyper-K  1$\sigma$ measurement uncertainty (in degrees) on
$\delta_{CP}$ for
$\sin^{2}\theta_{13} = 0.10$ assuming the baseline 10 year data sample with a 560~kton Gd-doped water detector.  (Statistical only errors are shown in parentheses.) Also shown is an estimate for the JPARC@Hyper-K sensitivity for a 560 kton water detector run for 7.5~MWyrs (3~years $\nu$ and 7 years~$\bar\nu$) assuming 5\% systematic errors and the LBNE experiment with a 35~kton liquid argon detector run for 8.5~MWyrs (5~years $\nu$ and 5~years $\bar\nu$).   }\label{deltaCP_123_127}%
\end{table}%

The DAE$\delta$ALUS high-statistics antineutrino data can be combined
with a neutrino-only long baseline measurement to provide improved
sensitivity for measuring $\delta_{CP}$.   One possibility is a
ten~year neutrino-only run of the JPARC@Hyper-K configuration combined with a
ten~year DAE$\delta$ALUS@Hyper-K exposure.  The complementarity of the two
experiments allows for a very precise search for $CP$ violation with
uncertainties estimated to be around $5^\circ$.      For this
discussion we make the same assumptions are were used for 
Table~\ref{events_123_127_HK}:  a 560 kton Gd-doped water detector,
 $\sin^2 2\theta_{13}=0.1$, and $\theta_{23}=49^\circ$ \cite{GonzalezGarcia:2012sz}.

The power 
of the combined run is shown in Figure~\ref{cp_error}. The top 
plot shows the expectation for the two experiments individually.
Nominal JPARC@Hyper-K  running assumes three years of running in neutrino mode.
This data set would yield the uncertainty indicates by the green
diamonds.   This would be followed by seven years of running in
antineutrino mode.   This data set, alone, results in the curve
indicated by the green $\times$ symbols.     One clearly sees that the
strength of JPARC@Hyper-K  is in neutrino running, as one would expect from a
conventional neutrino beam.    Combining these two data sets gives the
green solid curve with triangles.    
DAE$\delta$ALUS@Hyper-K alone, with a 10
year run, results in the solid red curve.   One can see
that DAE$\delta$ALUS@Hyper-K has a similar shape to the 
JPARC@Hyper-K antineutrino
running, where the differences come from the additional purity of the $\bar
\nu_e$ flux and the lack of a mass hierarchy effect in
the antineutrino data sample from DAE$\delta$ALUS.    We are proposing to combine a 
JPARC@Hyper-K run in
neutrino-mode only with the DAE$\delta$ALUS@Hyper-K pure antineutrino data 
set.      The result is shown on Figure~\ref{cp_error}, bottom, by the
black curve.   The individual contributions of the experiments are  
also shown.   One can see the complementarity, where DAE$\delta$ALUS@Hyper-K
provides the strong reach for $\delta_{cp}<0$ and JPARC@Hyper-K  provides the 
strong reach for $\delta_{cp}>0$ 

\begin{figure}[htbp]
\begin{center}
{\includegraphics[width=5.in]{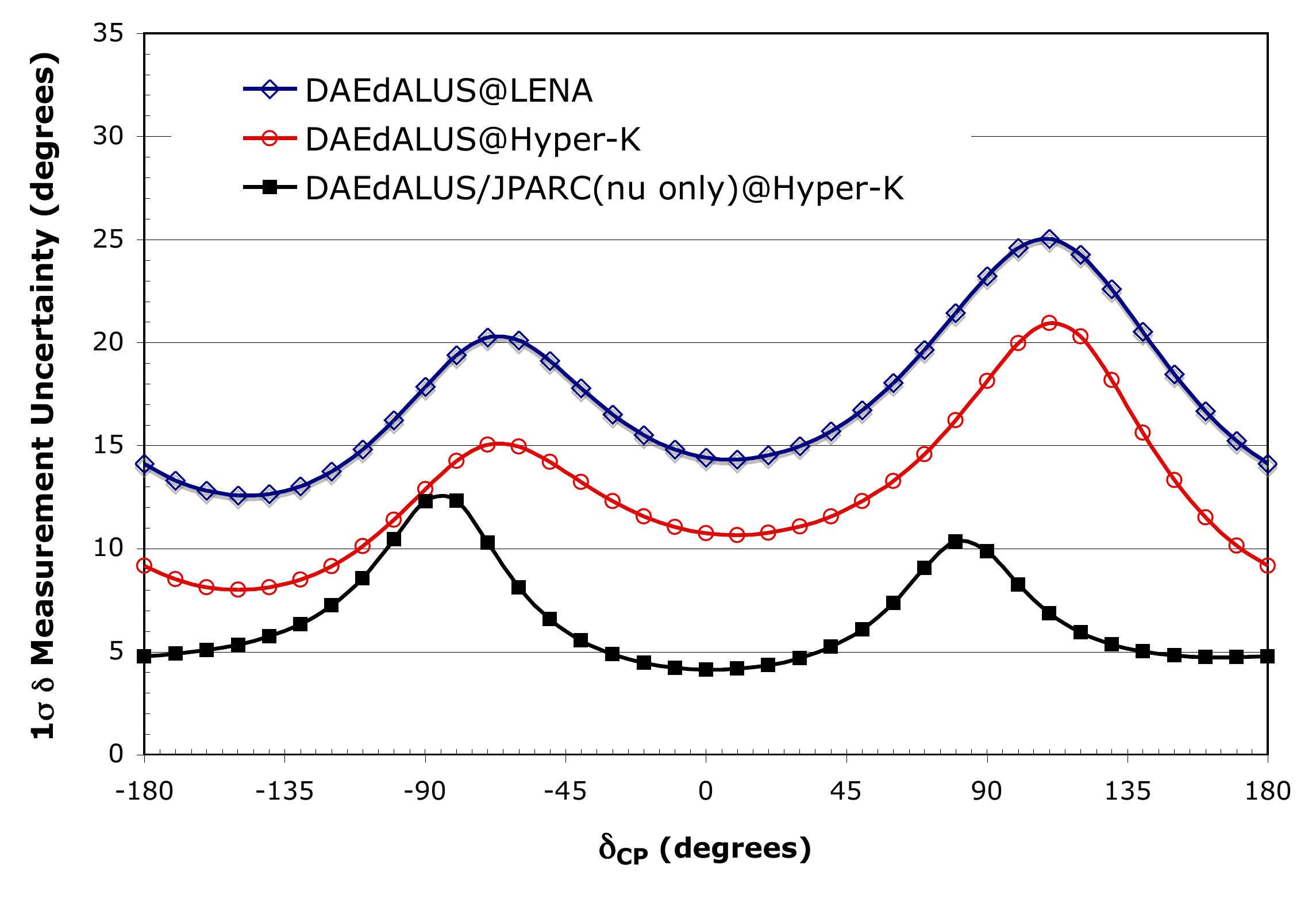}}\\
{\includegraphics[width=5.in]{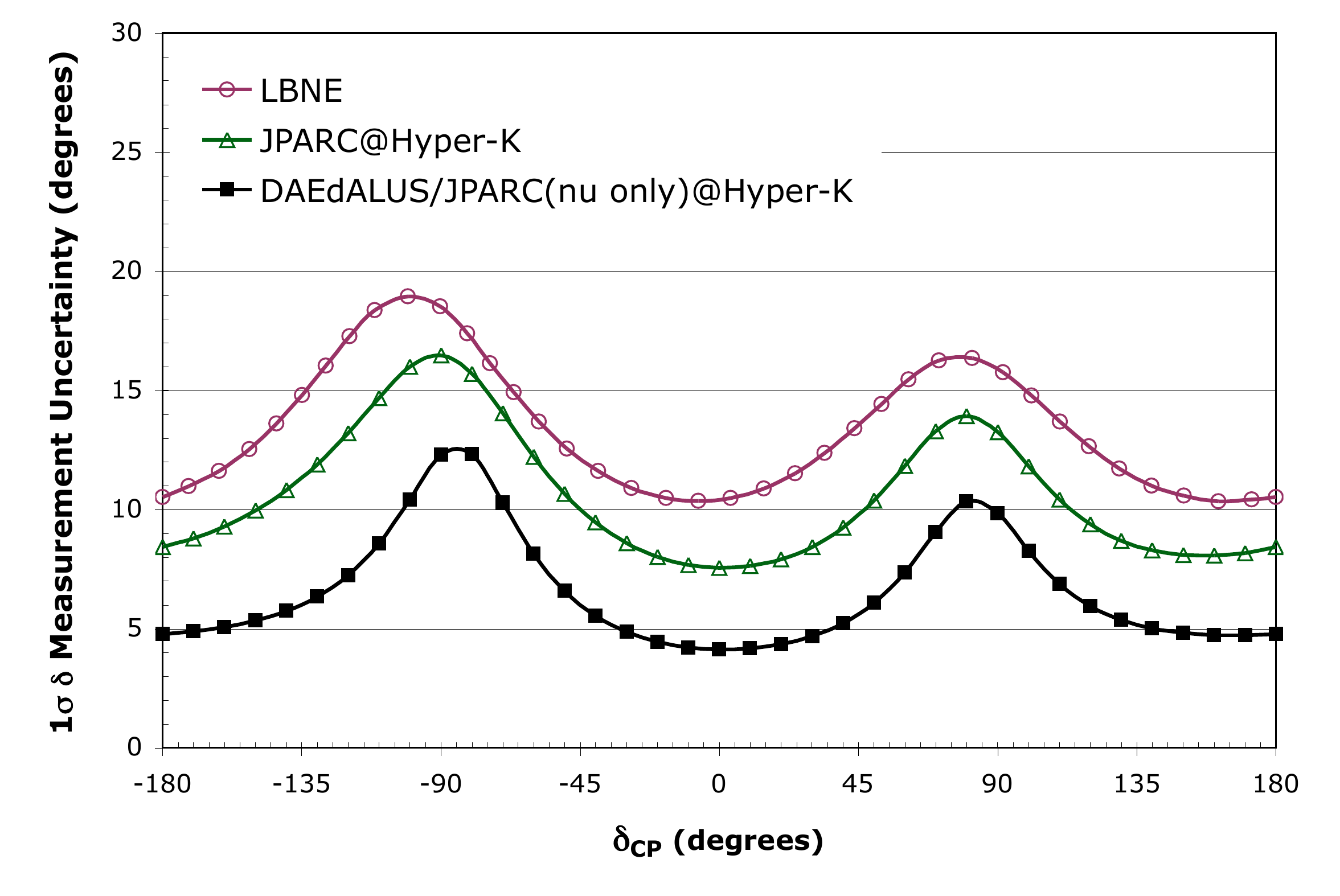}}
\end{center}
\caption{ Top:
The sensitivity of the $CP$-violation search in various
configurations:  Dark Blue -- DAE$\delta$ALUS@LENA,
Red-DAE$\delta$ALUS@Hyper-K,
Black--DAE$\delta$ALUS/JPARC(nu-only)@Hyper-K.
Bottom:   Light Blue-- LBNE;  Green-- JPARC@Hyper-K \cite{hyperk}
Black--DAE$\delta$ALUS/JPARC(nu-only)@Hyper-K
(same as above).   See Table~\ref{tab:configs} for the description of each
configuration.    
\label{Dsense}}
\end{figure}

\begin{figure}[htbp]
\begin{center}
{\includegraphics[width=5.in]{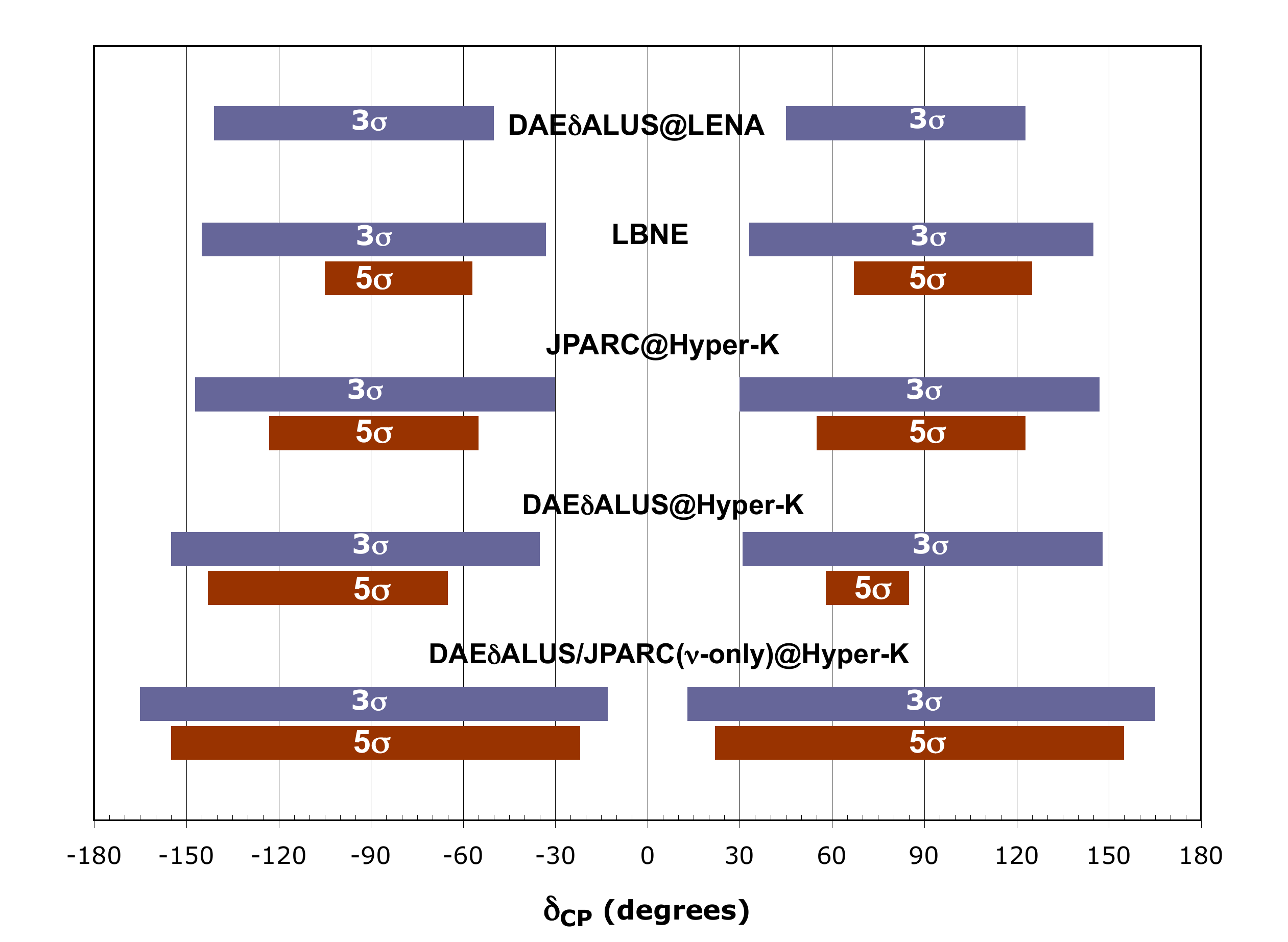}}
\end{center}
\caption{The $\delta_{CP}$ regions where an experiment can 
discover $CP$ violation by excluding $\delta_{CP} = 0^\circ$ or $180^\circ$ at 
3$\sigma$ or 5$\sigma$.
See Table~\ref{tab:configs} for the description of each
configuration.    
\label{comp_bars}}
\end{figure}

Finally, Figure~\ref{Dsense} shows the cross comparison of the experimental
configurations shown in Table~\ref{tab:configs}.   
The the combination DAE$\delta$ALUS/JPARC(nu-only)@Hyper-K configuration 
is compared to the two DAE$\delta$ALUS only configurations
using the LENA and Hyper-K detectors in Figure~\ref{Dsense} (top) and to 
JPARC@Hyper-K and LBNE in Figure~\ref{Dsense} (bottom).
From this figure, it is clear that the
DAE$\delta$ALUS/JPARC(nu-only)@Hyper-K configuration has impressive 
sensitivity to $\delta_{CP}$ with a significantly smaller measurement error as
compared to any of the other scenarios.  Figure~\ref{comp_bars} shows a comparison
of the $\delta_{CP}$ regions where an experiment can 
discover $CP$ violation by excluding the $\delta_{CP} = 0^\circ$ or $180^\circ$ at 
3$\sigma$ or 5$\sigma$.  Again, the DAE$\delta$ALUS/JPARC(nu-only)@Hyper-K
experiment clearly has substantially better coverage.

\subsection{Other physics with DAE$\delta$ALUS}
A number of other potential physics opportunities complement the main goal of a measurement of the neutrino $CP$
violating phase with DAE$\delta$ALUS.   These experiments can be
located near any of the three cyclotron locations.    We provide three examples here.    

A large water detector, used for a $\sin^2 \theta_W$ measurement
using $\nu_e$-electron scattering \cite{AgarwallaWeak},  would be complementary to the $\bar
\nu_e$-electron search in two ways.  First, it explores  differences in neutrinos versus antineutrinos that can be introduced
by new physics processes.   Second, with the ultra-high statistics of
the DAE$\delta$ALUS machines, an energy-dependent analysis, rather
than a rate analysis, becomes possible.     

A search for $\nu_\mu \rightarrow \nu_e$ appearance at high $\Delta
m^2$, hence short-baseline, can be performed to address the
LSND\cite{LSNDOsc} and MiniBooNE \cite{Aguilar-Arevalo:2013pmq} signals,  if the
DAE$\delta$ALUS configuration uses a large scintillator detector such as LENA \cite{agarwallaconradshaevitz}. 
In this case, the cyclotron must be located underground, within  tens of meters of the detector.
Like the $\nu_e$ disappearance search of IsoDAR, this study searches for the oscillation wave across the detector.     
Thus it would be powerful confirmation of a sterile-neutrino-related 
oscillation signal as the source of this high $\Delta m^2$ anomaly.
The $\nu_e$ flux can also be used for oscillation studies via the disappearance channel~\cite{agarwallaconradshaevitz}.

Lastly, a discovery of coherent neutrino scattering is also possible at a DAE$\delta$ALUS cyclotron. Notably, a cyclotron can provide the source of neutrinos for a coherent discovery at a deep underground detector 1.5~km away from the source~\cite{Anderson:2011bi} with only a small effect on such a detector's dark matter search exposure.  Furthermore, hundreds of coherent events can be collected with a dark-matter-style detector close to such a source for non-standard neutrino interaction sensitivity. A sensitive, unique neutral-current-based sterile neutrino search using coherent events can also be accomplished~\cite{Anderson:2012pn}.

\section{Conclusion}

At the 100th anniversary of Pontecorvo's birth,  neutrino physics is
entering a new ``precision era.''     To achieve our goals for the
next 100 years, 
improved flux sources are needed.  Decay-at-rest sources, driven by 
cyclotron accelerators, offer neutrino beams of 
well-defined flavor content and with energies in ranges where
backgrounds are low and knowledge of the cross section is high.   
This paper describes schemes to produce isotope and pion/muon 
decay-at-rest sources, developed as a part of the 
DAE$\delta$ALUS program.   

This paper has provided examples of the value of the high
precision beams for pursuing new physics.   In
particular, new results on a combined DAE$\delta$ALUS--Hyper-K search for
$CP$-violation are presented.     This study shows that errors on the mixing matrix parameter
ranging from 4\% to 12\% are achievable.   While this result is a
centerpiece of the program,   DAE$\delta$ALUS, and its early phase
program, IsoDAR,  allow for a wide range of important measurements and
searches.    Many examples have been presented here, focussing
primarily on searches for beyond Standard Model Physics through 
oscillations and non-standard interactions.     This establishment of
this rich new program is an great way for today's neutrino physicists--the intellectual descendants 
of Pontecorvo--to celebrate the anniversary of his birth.

\begin{center}
{ {\bf Acknowledgments}}
\end{center}
The authors thank the DAE$\delta$ALUS collaboration for useful
discussions.  Some studies reported here were initiated at the
2012 Erice International School of Subnuclear Physics Workshop,
supported by the Majorana Centre from the INFN Eloisatron Project,
which is directed by Prof. Antonino Zichichi.  The authors thank
the National Science Foundation for support. LW is supported by funds from UCLA.  

\bibliographystyle{apsrev}
\bibliography{cyclonu}

\end{document}